\documentclass{article}

\usepackage{arxiv}

\usepackage[utf8]{inputenc} 
\usepackage[T1]{fontenc}    
\usepackage[numbers]{natbib}
\usepackage{amsmath}
\usepackage{mhchem}
\usepackage{graphicx}
\usepackage{hyperref}       
\usepackage{url}            
\usepackage{booktabs}       
\usepackage{amsfonts}       
\usepackage{nicefrac}       
\usepackage{microtype}      
\usepackage{lipsum}

\title{Phase equilibrium of water with hexagonal and cubic ice using the scan functional}

\author{
    Pablo M. Piaggi \\
  Department of Chemistry, Princeton University, Princeton, NJ 08544, USA \\
  \texttt{ppiaggi@princeton.edu}
  \And
  Athanassios Z. Panagiotopoulos \\
  Department of Chemical and Biological Engineering, Princeton University, Princeton, NJ 08544, USA \\
  Princeton Institute for the Science and Technology of Materials, Princeton University, Princeton, NJ 08544, USA \\
  \And
  Pablo G. Debenedetti \\
  Department of Chemical and Biological Engineering, Princeton University, Princeton, NJ 08544, USA \\
  Princeton Institute for the Science and Technology of Materials, Princeton University, Princeton, NJ 08544, USA \\
  \And
  Roberto Car \\
  Department of Chemistry, Princeton University, Princeton, NJ 08544, USA \\
  Princeton Institute for the Science and Technology of Materials, Princeton University, Princeton, NJ 08544, USA \\
  Department of Physics, Princeton University, Princeton, NJ 08544, USA\\
  Program in Applied and Computational Mathematics, Princeton University, Princeton, NJ 08544, USA \\
}

\begin{document}
\maketitle

\begin{abstract}
Machine learning models are rapidly becoming widely used to simulate complex physicochemical phenomena with \textit{ab initio} accuracy. 
Here, we use one such model as well as direct density functional theory (DFT) calculations to investigate the phase equilibrium of water, hexagonal ice (Ih), and cubic ice (Ic), with an eye towards studying ice nucleation. 
The machine learning model is based on deep neural networks and has been trained on DFT data obtained using the SCAN exchange and correlation functional. 
We use this model to drive enhanced sampling simulations aimed at calculating a number of complex properties that are out of reach of DFT-driven simulations and then employ an appropriate reweighting procedure to compute the corresponding properties for the SCAN functional.
This approach allows us to calculate the melting temperature of both ice polymorphs, the driving force for nucleation, the heat of fusion, the densities at the melting temperature, the relative stability of ice Ih and Ic, and other properties.
We find a correct qualitative prediction of all properties of interest.
In some cases, quantitative agreement with experiment is better than for state-of-the-art semiempirical potentials for water.
Our results also show that SCAN correctly predicts that ice Ih is more stable than ice Ic.
\end{abstract}


\section{Introduction}

The transformation of liquid water into the beautifully ordered patterns of ice has long fascinated humans. The reason for this interest, however, has shifted over the years. Before the scientific revolution it might have stemmed from the aesthetic pleasure of observing snowflakes, or the fact that water and ice shape some of the most spectacular landscapes of planet Earth. Nowadays, ice and liquid water interest scientists due to their many anomalies such as the fact that ice is less dense than liquid water, the existence of a density maximum in the liquid around 4 $^o$C, the residual entropy of ice at 0 K, and the possible existence of a liquid-liquid transition at deeply supercooled conditions\cite{Debenedetti03}. Furthermore, from the point of view of technology, understanding the formation of ice is important in applications as diverse as weather prediction, cryopreservation, and food processing.

Ice nucleation refers to the initial formation of a microscopic ice cluster in liquid water. This process takes place over small time and length scales, and molecular simulations have long been recognized as an ideal tool to obtain insight into this phenomenon\cite{Matsumoto02}. Different semiempirical water models have been employed to study nucleation, ranging from the coarse-grained monoatomic water mW\cite{Molinero09}, to the realistic 4-site model TIP4P/Ice\cite{Abascal05} (see refs.~\citenum{Sanz13,Espinosa14,Li11,Yan12,Bauerecker08,Quigley08} for additional examples). These models are obtained by fitting their parameters to experimentally measured properties of water. Even though such models have been central to recent progress in many areas of water research\cite{Debenedetti20,Sanz13}, it has become clear that they have limitations\cite{Vega11}.
On the one hand, the mW model gives a coarse grained description in which the H atoms are eliminated, and is thus unable to capture effects such as the entropic contribution from proton disorder in ice Ih and Ic.
On the other hand, TIP4P/Ice considers rigid molecules based on the energy scale separation between intramolecular covalent bonding and intermolecular hydrogen bonding. However, in this way the interplay between intra and intermolecular interactions is lost.

Another avenue to construct a water model is using \textit{ab initio} energies and forces derived from electronic structure calculations.
One of the most popular approaches in this context is Kohn-Sham density-functional theory (DFT) \cite{Kohn65} based on approximations to the exact exchange and correlation (XC) functional\cite{KohanoffBook,Gillan16}.
The semilocal Perdew-Burke-Ernzerhof (PBE) approximation\cite{Perdew96} for the XC functional has been very successful for condensed phase systems but nonetheless fails at capturing even qualitatively some important properties of liquid water and ice (for a review see ref.\ \citenum{Gillan16}).
For instance, PBE is known to overstructure water\cite{Schmidt09,Gaiduk15} and to give a higher density for ice Ih than for water\cite{Chen17}.
These issues can be mitigated by including a fraction of exact exchange and adding van der Waals interactions\cite{Lin12,Gillan16,Gaiduk15,Cheng19}. Recently, the strongly constrained and appropriately normed (SCAN) semilocal XC functional\cite{Sun15} has been proposed and it has been shown to provide a correct qualitative picture of liquid water and ice\cite{Sun16,Chen17} at an affordable computational cost. SCAN is non empirical and thus could provide a truly predictive description of water and ice.

In this work, we assess the fitness of the SCAN XC functional to describe the thermodynamics of liquid water, hexagonal ice, and cubic ice.
The knowledge provided by this investigation will be important in future studies of ice nucleation. 
Since the calculation of many properties is out of reach of DFT-driven molecular dynamics (MD) simulations\cite{Car85}, here we employ a deep neural network potential (NNP) trained on SCAN DFT data in combination with enhanced sampling methods.
In this way, we are able to access the time and length scales needed to calculate the relevant properties.
We then use an appropriate reweighting procedure to obtain the corresponding properties for the SCAN model itself.
This procedure provides a fine validation of the NNP. 
We report on the melting temperature of ice Ih and ice Ic, the heat of fusion of ice Ih, the driving force for nucleation (supersaturation),  the relative stability of ice Ih and Ic, and other properties.

The calculations described herein neglect nuclear quantum effects (NQE).
An experimental manifestation of NQE are the observed isotope shifts of various thermodynamic properties\cite{Soper08}.
In principle these effects could be taken into account by path integral MD simulations, which however would increase significantly the cost of the simulations.
Since the isotope shifts are small (up to several K) it is reasonable to neglect them as a first approximation.
While NQE on properties like the melting temperature have been studied recently using machine learning models\cite{Cheng19,Yao20}, including their effect on nucleation studies is still computationally impractical.

\section{Methods}
\label{sec:methods}

The route that we  use to calculate properties of water and ice is based on three steps.
First, we run a molecular dynamics simulation driven by the NNP model introduced in ref.\ \citenum{Gartner20}.
At this stage we make extensive use of enhanced sampling methods.
We then extract configurations from these simulations and  calculate the energy using DFT based on the SCAN XC functional.
Lastly, we employ an appropriate reweighting procedure to obtain ensemble averages at the SCAN level.
In the following sections we describe in detail each of these steps.

\subsection{MD simulations driven by the NNP model}
Molecular dynamics simulations were performed using LAMMPS\cite{Plimpton95} interfaced with the DeePMD-kit\cite{Wang18}.
We employed the deep NNP model developed in ref.\ \citenum{Gartner20} that is based on the DeePMD framework proposed by Zhang et al\cite{Zhang18,Zhang18end}.
This potential was trained with the DP-GEN software\cite{Zhang20} that uses an active learning approach\cite{Zhang19}.
The training data set was obtained with the SCAN XC functional and spans a large region of water's phase diagram including configurations of liquid water and 15 ice polymorphs.
A kinetic energy cutoff of 110 Ry was used for the wavefunctions during the training.
Further details of the training procedure can be found in ref.~\citenum{Gartner20}.
This potential was recently used to provide first principles evidence consistent with the existence of a second critical point in deeply supercooled water\cite{Gartner20}.
Furthermore, we note that a highly optimized version of DeePMD on massive GPU platforms has been shown to be able to simulate efficiently several millions of water molecules\cite{Lu20} opening the door to study complex phenomena, such as ice nucleation, with \textit{ab initio} accuracy.

A time step of 0.5 fs was used in all simulations.
Temperature was kept constant using the stochastic velocity-rescaling thermostat\cite{Bussi07} with a relaxation time of 0.1 ps.
Pressure was maintained at 1 bar using a Parrinello-Rahman barostat\cite{Parrinello81} with a relaxation time of 1 ps.
We performed liquid water-ice Ih coexistence simulations using a system of 576 molecules.
In this type of simulation a crystal of ice Ih containing proton disorder is brought into contact with the liquid phase.
Configurations of ice Ih with proton disorder were obtained using the software GenIce\cite{Matsumoto18}.
The secondary prismatic plane ($1\bar{2}10$) of ice Ih was exposed to the liquid since it is the fastest growing interface of ice Ih\cite{Conde17}.
The sides of the box perpendicular to the interface were set to the NNP equilibrium size of the corresponding ice Ih crystal (see ESI for details), and kept at that size throughout the simulation.
The side of the box parallel to the interface was controlled with a Parrinello-Rahman barostat set at 1 bar.
Four simulations were performed at each temperature with different random seeds for the initial velocities.
We also employed 576 molecules for the liquid water-ice Ic coexistence simulations.
In this case, the (100) face of ice Ic was exposed to the liquid and the initial crystal also contained proton disorder.
The procedure to determine the sides of the simulation box are analogous to that used for ice Ih and the same barostat setup was employed.

\subsection{Enhanced sampling simulations}

Enhanced sampling simulations are based on the introduction of a bias potential $V(\mathbf{s})$ along a suitable set of collective variables (CVs) $\mathbf{s}$. The CVs are continuous and differentiable functions of the atomic coordinates $\mathbf{R}$ and possibly the volume $\mathcal{V}$. 
The bias potential alters the marginal probability distribution of $\mathbf{s}$ that in the isothermal-isobaric ensemble is,
\begin{equation}
    p(\mathbf{s})=\int d\mathbf{R} \: d\mathcal{V} \frac{e^{-\beta [U(\mathbf{R})+P\mathcal{V}]}}{Z_{\beta,P}} \delta(\mathbf{s}-\mathbf{s}(\mathbf{R})),
\end{equation}
where $U(\mathbf{R})$ is the potential energy, $Z_{\beta,P}$ is the appropriate partition function at inverse temperature $\beta = 1/(k_B T)$ and pressure $P$, $T$ is the temperature, and $k_B$ is the Boltzmann constant.
If the desired target marginal probability distribution $p^{\mathrm{tg}}(\mathbf{s})$ is known then the bias potential can be defined as,
\begin{equation}
    V(\mathbf{s})=-\frac{1}{\beta}\log\left ( \frac{p^{\mathrm{tg}}(\mathbf{s})}{p(\mathbf{s})}\right ).
\end{equation}
Since $p(\mathbf{s})$ is generally not known \textit{a priori},  iterative methods have been proposed to determine the appropriate bias potential\cite{Laio02,Barducci08,Valsson14,Invernizzi20rethinking}.

Enhanced sampling methods are often used to study rare events by choosing a $p^{\mathrm{tg}}(\mathbf{s})$ in which the probability of observing transition state configurations is greatly increased.
Here, we use enhanced sampling to sample a multithermal ensemble and to simulate the crystallization of ice Ih and Ic. 
In order to perform these simulations we augmented LAMMPS with the PLUMED enhanced sampling plugin\cite{Tribello14,Bonomi19}.

\subsubsection{Multithermal simulations}

In the isothermal-isobaric ensemble only a relatively small energy range is sampled with high probability.
By contrast, in the multithermal-isobaric ensemble a much larger energy range is explored with high probability in order to obtain information about the system at all temperatures in some predefined temperature interval $[T_1,T_2]$.
Recently, it has been shown that multithermal-multibaric ensembles can be sampled using collective-variables-based methods and taking the potential energy and the volume as collective variables\cite{Piaggi19}.
Here we shall use the on-the-fly probability enhanced sampling (OPES) method\cite{Invernizzi20rethinking,Invernizzi20}, and the potential energy and the volume as CVs in order to sample a multithermal ensemble.
This procedure has the advantage that from a single simulation one can calculate the ensemble average $ \langle O(\mathbf{R},\mathcal{V}) \rangle$ of any observable $O$ that is a function of $\mathbf{R}$ and $\mathcal{V}$ at any temperature $T$ provided that $T_1 \leq T \leq T_2$.
Since the simulation is performed in a biased ensemble we use the following formula to obtain ensemble averages in the isothermal-isobaric ensemble,
\begin{equation}
\langle O(\mathbf{R},\mathcal{V}) \rangle_{\beta'} = \frac{ \langle O(\mathbf{R},\mathcal{V}) w(\mathbf{R},\mathcal{V}) \rangle_{\beta,V}}
                                                      { \langle w(\mathbf{R},\mathcal{V}) \rangle_{\beta,V}},
\label{eq:reweight}
\end{equation}
where $w(\mathbf{R},\mathcal{V})=e^{(\beta-\beta')U(\mathbf{R})} e^{\beta V}$, $\langle \cdot \rangle_{\beta'}$ is the ensemble average in the isothermal-isobaric ensemble at inverse temperature $\beta'$ and some pressure $P$, $\langle \cdot \rangle_{\beta,V}$ is the ensemble average at inverse temperature $\beta$ and pressure $P$ with bias potential $V$.
We performed several multithermal simulations and the details are provided in the ESI.
Properties of ices were averaged over different proton configurations obtained with the software GenIce\cite{Matsumoto18}.
An isotropic barostat was used for liquid water and a fully anisotropic barostat was used for ices.
The enthalpy and density were calculated from these simulations using Eq.\ \eqref{eq:reweight}.
All errors were calculated using weighted block averages as described in ref.~\citenum{Invernizzi20}.
The error contribution from the finite sampling of proton configurations was taken into account as discussed in the Appendix.

\subsubsection{Crystallization simulations.}
\label{subsubsec:crystallization}

We are interested in calculating chemical potentials and melting temperatures of ices.
The chemical potential difference between ice and liquid water can be obtained from the expression:
\begin{equation}
\Delta\mu_{l \rightarrow ice} (T) = -\frac{1}{\beta N} \log\left ( \frac{p_{ice}}{1-p_{ice}}\right )
\label{eq:chem_pot_liquid_ice_population}
\end{equation}
where $p_{ice}$ is the probability of finding the system in the solid phase and $N$ is the number of molecules in the system.
Many transitions between water and ice must be observed in order to calculate $p_{ice}$ accurately.
Since crystallization is a first order phase transition, a free energy barrier must be surmounted to proceed with the transformation.
In a standard molecular dynamics simulation the probability of observing the transition is extremely low and $p_{ice}$ cannot be estimated.
Enhanced sampling methods allow to circumvent this problem by increasing the probability of crossing the free energy barrier.

Here we simulate the crystallization process using enhanced sampling and a CV tailored to describe the crystal structures of ice Ih and Ic.
These simulations are similar to those of ref.~\citenum{Piaggi20} that were used to calculate the difference in chemical potential between liquid water and ice Ih, and the melting temperature of ice Ih in the TIP4P/Ice model of water.
We use the Environment Similarity CV \cite{Piaggi19b} that counts the number of atomic environments in the simulation that are compatible with a reference environment.
We define two CVs:
The first, $s_{Ih}$, counts the number of atomic environments compatible with the environments of ice Ih that exist in the simulation box\cite{Piaggi20}.
We only consider oxygen atoms since hydrogen atoms exhibit disorder in ice Ih and therefore a particular environment should not be enforced.
There are four reference environments in ice Ih and each reference environment contains 17 neighbors.
The second CV, $s_{Ic}$, counts the number of atomic environments compatible with the environments of ice Ic.
In this case there are only two reference environments each with 16 neighbors.
Both CVs go from zero in the liquid phase to $N$ in the solid phase. 

Two types of crystallization simulations were performed.
The first type of simulations aimed at exploring a single temperature and were performed using the variational principle of Valsson and Parrinello\cite{Valsson14} and the VES code\cite{vescode}.
The bias potential in this case is one-dimensional and aims at obtaining a uniform sampling of the chosen CV.
The second type of simulations were performed in a generalized multithermal ensemble\cite{Piaggi19}.
These simulations use a variant of the OPES method that constructs a three-dimensional bias potential which is a function of the potential energy, the volume, and either $s_{Ih}$ or $s_{Ic}$.
The temperature interval 300-350 K was targeted in this case.

In both cases the introduction of a bias potential results in multiple reversible transitions between liquid water and ice such that ergodic sampling is achieved.
The simulations thus reproduce the crystallization process although small system sizes are used and therefore a critical nucleus is never observed.
Proton configurations of the ices are also sampled properly since each new crystallization generates a new proton configuration\cite{Piaggi20}.
Separate simulations were used to crystallize ice Ih and Ic.
The crystallization of ice Ih is driven by $s_{Ih}$ and that of ice Ic is driven by $s_{Ic}$. 
The size and shape of the simulation box was chosen to accommodate perfect ice crystals.
Furthermore, $s_{Ih}$ and $s_{Ic}$ are non rotationally invariant and therefore enforce a particular orientation of the crystal structure.
This is important to avoid crystals with defects or strains.
An isotropic barostat was used to maintain constant atmospheric pressure and to avoid changing the shape of the box when the system is in the liquid state.
A list of the simulations that were performed and the details to reproduce them are provided in the ESI.
The evolution of $s_{Ih}$ or $s_{Ic}$ as a function of simulation time is also shown in the ESI.

From these simulations $p_{ice}$ can be computed using $p_{ice}=\langle H(s-s_0)\rangle$ where $\langle \cdot\rangle$ denotes an ensemble average, $H$ is the unit step function, and $s_0$ is a threshold that separates liquid from ice configurations.
For large free energy barriers, the choice on $s_0$ is not crucial and a simple choice is $s_0=N/2$.
Since the simulations were performed with a bias potential, Eq.~\eqref{eq:reweight} has to be used to compute $p_{ice}=\langle H(s-s_0)\rangle$.
Errors $\sigma_{p_{ice}}$ in the calculation of $p_{ice}$ were computed using weighted block averages and propagated to the error in $\Delta\mu_{l \rightarrow ice}$ using the formula $\sigma_{\Delta\mu} = \sigma_{p_{ice}} / ( \beta N p_{ice}(1-p_{ice}) )$.

\subsection{DFT calculations}

We performed DFT calculations in order to validate the NNP and to calculate properties of the DFT model.
Configurations were extracted at regular intervals from the simulations driven by the NNP.
The simulations used and the stride between extracted configurations are described in the ESI.
Afterwards, SCAN DFT calculations were performed on these configurations using  the Quantum Espresso\cite{Giannozzi09,Giannozzi17} suite for electronic structure calculations.
The SCAN exchange and correlation functional was evaluated with the LIBXC 4.3.4 library\cite{Marques12}.
We employed norm-conserving, scalar-relativistic pseudopotentials for O and H parameterized using the PBE functional\cite{Hamann13}.
Kinetic energy cutoffs of 150 Ry and 600 Ry were used for the wavefunctions and the charge density, respectively.
These cutoffs were chosen in order to have an error of around 10 J/mol per molecule in the energy difference between cubic and hexagonal ice (see ESI for details).
We note that the wavefunction kinetic energy cutoff used here is larger than the one used to obtain the training data in ref.~\citenum{Gartner20}.
Only the $\Gamma$ point of the Brillouin zone was sampled and the convergence absolute error for the self-consistent procedure was set to $10^{-6}$ Ry.
For the calculation of the difference in enthalpy between ice Ih and ice Ic, the energy was minimized with respect to the atomic coordinates and the cell vectors.
The minimization ended when the change in energy was less than $10^{-4}$ Ry.
All other parameters were set to their default values in Quantum Espresso.

\subsection{Reweighting from the NNP to a DFT model}

Machine learning models are not perfect representations of the underlying DFT potential energy surface. 
Thus, it is of interest to understand the properties of the DFT model itself, rather than the properties of the approximate machine learning model. 
The direct calculation of complex DFT properties is often not possible due to the high cost of DFT-driven dynamics.
Here, we employ a different strategy based on a combination of dynamics driven by the machine learning model and a subsequent reweighting.

The procedure works as follows.
We consider an observable $O(\mathbf{R},\mathcal{V})$ that is a function of the coordinates $\mathbf{R}$ and the volume $\mathcal{V}$.
We also consider the potential energy $U^{NNP}(\mathbf{R},\mathcal{V})$ of the NNP and the potential energy $U^{DFT}(\mathbf{R},\mathcal{V})$ of the DFT model.
The ensemble average of $O(\mathbf{R},\mathcal{V})$ in the isothermal-isobaric ensemble at inverse temperature $\beta$ and pressure $P$ is,
\begin{equation}
\langle O(\mathbf{R},\mathcal{V}) \rangle_{\beta}^{NNP} = \int d\mathbf{R} \: d\mathcal{V} \:  O(\mathbf{R},\mathcal{V}) \frac{e^{-\beta [U^{NNP}(\mathbf{R},\mathcal{V})+P\mathcal{V}]}}{Z_{\beta}^{NNP}}
\label{eq:ensemble_average_nnp}
\end{equation}
in the NNP and,
\begin{equation}
\langle O(\mathbf{R},\mathcal{V}) \rangle_{\beta}^{DFT} = \int d\mathbf{R} \: d\mathcal{V} \: O(\mathbf{R},\mathcal{V}) \frac{e^{-\beta [U^{DFT}(\mathbf{R},\mathcal{V})+P\mathcal{V}]}}{Z_{\beta}^{DFT}}
\label{eq:ensemble_average_dft}
\end{equation}
in the DFT model where $Z_{\beta}^{NNP}$ and $Z_{\beta}^{DFT}$ are the appropriate partition functions.
We would typically calculate the ensemble averages in Eqs.~\eqref{eq:ensemble_average_nnp} and \eqref{eq:ensemble_average_dft} using dynamics driven by the NNP and the DFT model, respectively. However, using the equations above we can show that  $\langle O(\mathbf{R},\mathcal{V}) \rangle_{\beta'}^{DFT}$ at any temperature $\beta'$ can be written as,
\begin{equation}
\langle O(\mathbf{R},\mathcal{V}) \rangle_{\beta'}^{DFT} = \frac{\langle O(\mathbf{R},\mathcal{V}) e^{ \beta (U^{NNP}+V) -\beta'U^{DFT} } \rangle_{\beta,V}^{NNP}}{\langle  e^{ \beta (U^{NNP}+V) -\beta'U^{DFT} } \rangle_{\beta,V}^{NNP}}
\label{eq:rew_dft}
\end{equation}
where $\beta$ is the temperature at which the simulation was performed, and $V$ is the bias potential.
We note that the dependence of $U^{NNP}$, $U^{DFT}$, and $V$ on $\mathbf{R}$ and $\mathcal{V}$ has been dropped.
Eq.~\eqref{eq:rew_dft} shows that ensemble averages of the DFT model can be obtained from simulations driven with the NNP model.
We used this formula to calculate chemical potential differences, melting temperatures, densities, and other properties.
We note that the well-known free energy perturbation formula of Zwanzig\cite{Zwanzig54} is a special case of Eqn.~\eqref{eq:rew_dft}.
The procedure described here is similar to the one used in ref.~\citenum{Chehaibou19}.

\section{Results and discussion}

\subsection{Melting temperature and water-ice chemical potential differences}
\begin{figure}[t]
\centering
  \includegraphics[width=0.55\textwidth]{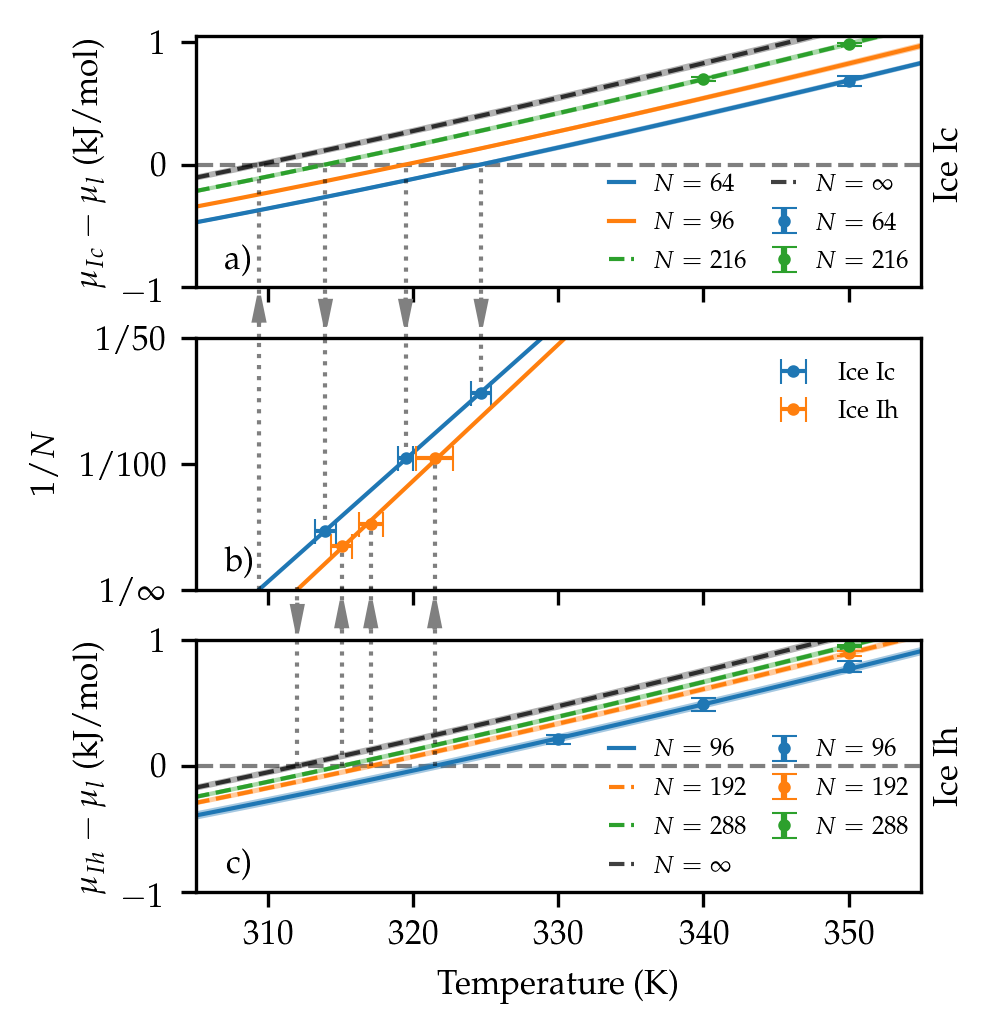}
  \caption{Results of free energy calculations for the NNP. 
  a) Difference in chemical potential as a function of temperature for ice Ic. Systems with $N=64$ , $96$, and $216$ molecules were employed. b) Melting temperature as a function of inverse system size ($1/N$). A straight line was fit to the results to extrapolate to the thermodynamic limit. c) Difference in chemical potential as a function of temperature for ice Ih. System sizes with $96$, $192$ and $288$ molecules were employed. The arrows between the plots indicate that the melting temperatures for each system size were used to calculate the melting temperature in the thermodynamic limit. Continuous lines correspond to multithermal crystallization simulations with the errors shown as shaded regions. Points with error bars are results from single temperature crystallization simulations. Dashed lines correspond to curves that have been shifted to extrapolate results of small system sizes to larger systems (see ESI for details of this procedure). All errors are calculated from four-fold block averages.}
  \label{fig:free_energy}
\end{figure}

The first property that we calculated was the melting temperature of ice Ih $T_m^{Ih}$ at atmospheric pressure for the NNP using the crystallization simulations described in Section \ref{subsubsec:crystallization}.
To this end, we calculated chemical potential differences between liquid water and each ice polymorph using Eq.~\eqref{eq:chem_pot_liquid_ice_population}.
The chemical potential difference between ice Ic and liquid water $\Delta\mu_{l \rightarrow Ic} (T)$ is shown in Figure \ref{fig:free_energy}a for systems with 64, 96, and 216 water molecules. 
An equivalent plot for the difference in chemical potential between ice Ih and liquid water $\Delta\mu_{l \rightarrow Ih} (T)$ is shown in Figure \ref{fig:free_energy}c for systems with 96, 192, and 288 molecules.
Details of the simulations are provided in the ESI.
From the chemical potentials, we calculated the melting temperature of each polymorph and system size using the fact that $\Delta\mu_{l \rightarrow Ih} (T_m^{Ih})=0$ and $\Delta\mu_{l \rightarrow Ic} (T_m^{Ic})=0$ where $T_m^{Ih}$ and $T_m^{Ic}$ are the melting temperatures of ice Ih and Ic.
The melting temperatures of both ices calculated in this fashion as a function of inverse system size are shown in Figure \ref{fig:free_energy}b.
Since finite size scaling theory for first order phase transitions predicts a linear relationship between the melting temperature and the inverse of the number of molecules\cite{Binder87}, we fit a straight line to the data in Figure \ref{fig:free_energy}b and obtain a melting temperature of 312 K for ice Ih and 309 K for ice Ic in the thermodynamic limit.
Using these melting temperatures, we shifted the chemical potential differences obtained for the finite sized systems by the corresponding amount; results are shown in black in Figures \ref{fig:free_energy}a and c.

\begin{figure}[t]
\centering
  \includegraphics[width=0.55\textwidth]{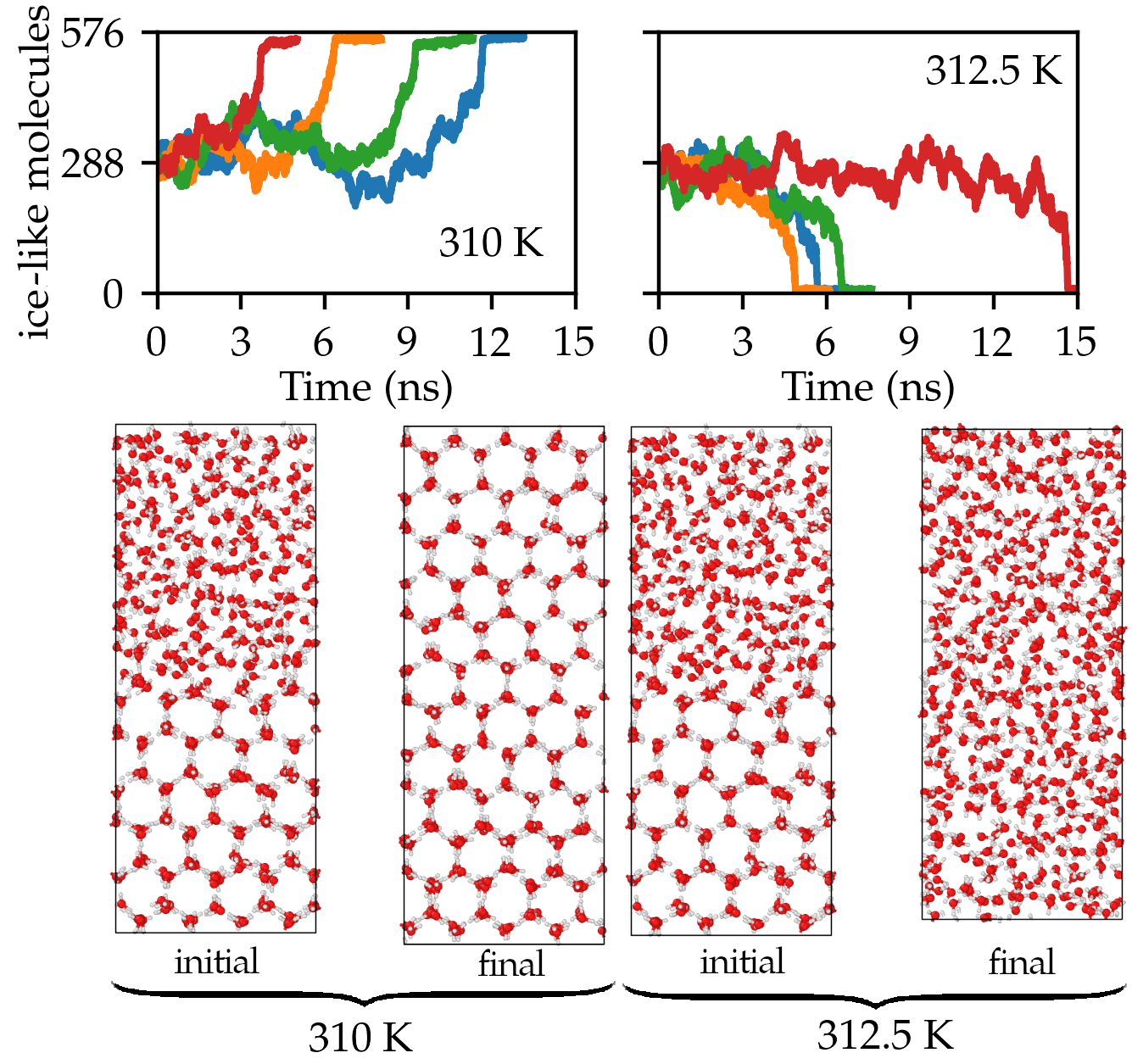}
  \caption{Direct coexistence simulations of liquid water and ice Ih. The number of ice Ih-like molecules\cite{Piaggi20} as a function of simulation time is shown. Four independent runs with different initial seeds for the velocities are shown in different colors at temperatures 310 K and 312.5 K. Representative initial and final configurations are shown with oxygen in red and hydrogen in white.}
  \label{fig:coexistence}
\end{figure}

We confirmed the melting temperatures obtained above for the NNP model by performing direct coexistence simulations.
The results of four simulations with different random seeds for the initial velocities at 310 K and 312.5 K are shown in Figure \ref{fig:coexistence} for the system liquid water-ice Ih. 
At 310 K the ice crystal grows at the expense of liquid water and at 312.5 K ice thaws.
It follows that the melting temperature of ice Ih for this model lies in the interval 310 K $<T_m^{Ih}<$ 312.5 K. 
We verified the protocol used for the coexistence simulations by calculating the melting temperature of ice Ih within the TIP4P/Ice\cite{Abascal05} model. These calculations are reported in the ESI and a melting temperature of 269 K is found in good agreement with the widely accepted value 270 K\cite{Conde17,Piaggi20}. 
We also employed coexistence simulations to calculate the melting temperature of ice Ic ($T_m^{Ic}$).
We found that at 307.5 K the crystal grows and that at 310 K the crystal melts.
Details of these simulations are discussed in the ESI.
We conclude that the melting temperature of ice Ic lies in the interval 307.5 K $<T_m^{Ic}<$ 310 K.

The melting temperatures from coexistence simulations are in excellent agreement with the estimates from the crystallization simulations.
However, the melting temperature of ice Ih is almost 40 K above the experimental melting temperature.
The lower melting temperature of ice Ic implies that for this model ice Ih is more stable than ice Ic at atmospheric pressure.
We will analyze the relative stability of ice Ih and Ic in greater detail below.
The melting temperatures obtained are summarized in Table \ref{tbl:melting} and also compared to experimental and TIP4P/Ice results.

\begin{figure}[t]
\centering
  \includegraphics[width=0.55\textwidth]{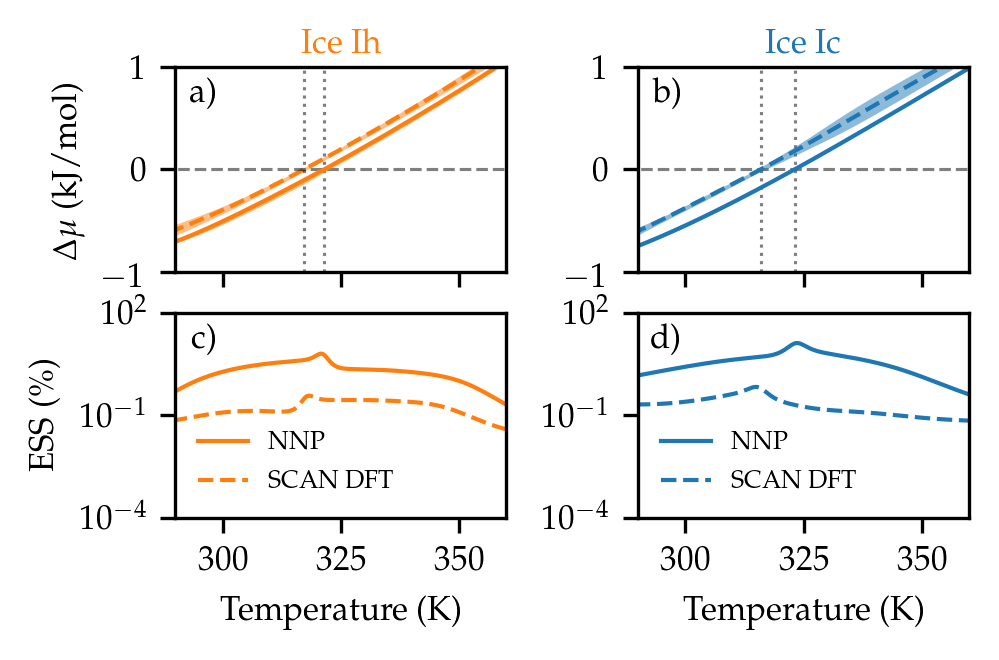}
  \caption{Calculation of chemical potential differences of SCAN DFT obtained through reweighting of the crystallization simulations driven by the NNP. a) and b) show the chemical potential difference between liquid water and ice Ih and ice Ic, respectively (full lines: NNP; dashed lines: SCAN DFT reweighting). Errors are shown as shaded regions and were calculated from two-fold and four-fold block averages for ice Ih and ice Ic, respectively. c) and d) show the effective sample size (ESS) in each case as a percentage of the total samples. This is a way to quantify the efficiency of the simulations for different temperatures and models. Systems of 96 and 64 molecules were used for ice Ih and ice Ic, respectively. Comparison of the results shown here with Table \ref{tbl:melting} shows appreciable size effects.}
  \label{fig:chem_pot_rew_scan}
\end{figure}

From the crystallization simulations one can also calculate $\Delta\mu_{l \rightarrow Ih}$ and $\Delta\mu_{l \rightarrow Ic}$ of the DFT model using the reweighting procedure in Eq.~\eqref{eq:rew_dft}.
In Figures \ref{fig:chem_pot_rew_scan} a and b we compare $\Delta\mu_{l \rightarrow Ih}$ and $\Delta\mu_{l \rightarrow Ic}$ calculated with the NNP and with SCAN DFT.
The analysis was performed on systems of 96 and 64 molecules for ice Ih and ice Ic, respectively.
We found that the melting temperatures are lowered by 4 K in ice Ih and by 7 K in ice Ic for SCAN DFT in comparison to the NNP. 
The differences between the NNP and SCAN DFT are partly due to the different cutoff for the SCAN calculations used to train the NNP and partly to inaccuracies of the NNP with respect to the underlying DFT energy surface.
Thus, the melting temperatures of ices in the thermodynamic limit for SCAN DFT are 308 K for ice Ih and 302 K for ice Ic.
These results are summarized in Table \ref{tbl:melting}.
This brings the melting temperature of ice Ih in better agreement with experiment but still 35 K above the experimental melting temperature.
The difference between the SCAN and experimental melting temperature could be attributed in part to the missing NQE and in part to the intrinsic accuracy of the SCAN functional approximation.
Indeed, Cheng et al\cite{Cheng19} found that NQE lowered the melting temperature by 8 K using a different DFT XC functional.

The reweighting of Eq.~\eqref{eq:rew_dft} can only work if there is overlap between the isothermal-isobaric distribution for DFT at a given temperature and the distribution sampled in the MD simulation.
Therefore if the NNP differs significantly from the DFT model the procedure will fail.
A useful way to quantify the efficiency of the reweighting procedure is the effective sample size (ESS).
The ESS is defined as $\mathrm{ESS}=(\sum w_i)^2/\sum w_i^2$ where $w_i$ is the unnormalized weight of the i-th configuration used for reweighting\cite{Kong94,Elvira18}.
We divide the ESS by the number of configurations used in order to obtain a sampling efficiency.
We show in Figure \ref{fig:chem_pot_rew_scan} c and d the ESS as a function of temperature both for the NNP model and DFT.
The efficiency of the multithermal crystallization simulations is relatively high for the NNP model at around 5 \%.
On the other hand the ESS falls to around 0.1 \% when the weights appropriate for DFT are considered.
This sampling efficiency is low but sufficient to obtain relatively small error bars and determine chemical potential differences with a reasonable accuracy.
This analysis highlights the importance of training an accurate machine learning model.

\begin{table}[b]
\small
  \caption{Melting temperature of ices}
  \label{tbl:melting}
  \centering
  \begin{tabular*}{0.48\textwidth}{@{\extracolsep{\fill}}ll}
    \hline
    \multicolumn{2}{c}{Ice Ih} \\
    \hline
     & Melting temperature (K) \\
    \hline
    Direct coexistence NNP & 310-312.5 \\
    Free energy calculations NNP & 312(1) \\
    Reweighting SCAN DFT & 308(2) \\
    TIP4P/Ice\cite{Piaggi20,Conde17} & 270  \\
    Experiment & 273.15 \\
    \hline
    \multicolumn{2}{c}{Ice Ic} \\
    \hline
     & Melting temperature (K) \\
    \hline
    Direct coexistence NNP & 307.5-310 \\
    Free energy calculations NNP & 309(1) \\
    Reweighting SCAN DFT & 302(2) \\
    \hline
  \end{tabular*}
\end{table}

\subsection{Validation of the NNP}

\begin{figure}[t]
\centering
  \includegraphics[width=0.55\textwidth]{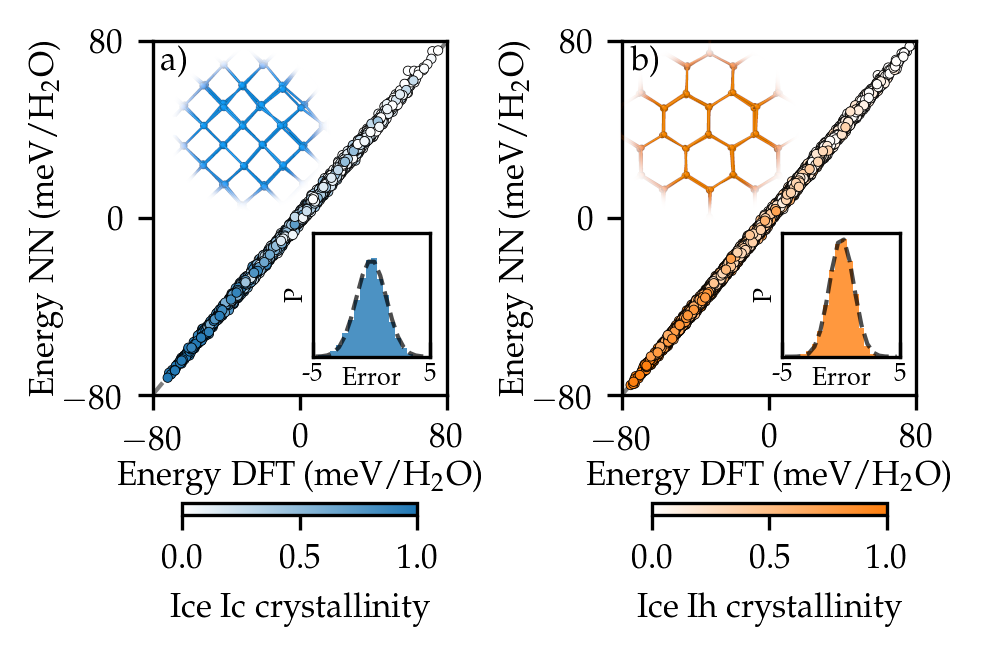}
  \caption{Validation of the NNP. Scatter plots of the NNP model energy vs the DFT energy. Atomic configurations were extracted from the simulations used to calculate melting temperatures. a) and b) correspond to simulations that reproduce the crystallization process for ice Ic and Ih, respectively. The insets shows the distribution of errors in meV/\ce{H2O} molecule. Dots in the scatter plot are colored according to the ice crystallinity that is the fraction of molecules that have environments compatible with ice Ic or ice Ic. }
  \label{fig:validation}
\end{figure}

The simulations used to calculate the melting temperatures reproduce the crystallization process and therefore are useful to validate the NNP for studying ice nucleation.
We therefore extracted configurations from a crystallization simulation of ice Ic with 64 molecules, and a simulation of ice Ih with 96 molecules.
These simulations were performed in the multithermal ensemble and contain configurations relevant to the temperature range 300-350 K.
For each of the extracted configurations we compared the potential energy of the NNP against energies calculated directly using DFT with the SCAN XC functional.
In Figure \ref{fig:validation} we show scatter plots of the NNP energy vs the DFT energy.
The dots are colored according to the ice crystallinity defined as the fraction of oxygen atoms that have environments compatible with ice Ic or ice Ih.
The crystallinity is computed from the $s_{Ih}$ and $s_{Ic}$ CVs described above.
The fact that the configurations span crystallinities from 0 to 1 means that we have chosen configurations representative of liquid water, ice, and also intermediate configurations representative of the crystallization process.
The insets in Figure \ref{fig:validation} show the distribution of the errors.
The agreement is excellent and the error distributions have deviations of 1.3 and 1.1 meV per \ce{H2O} molecule for the crystallization simulations of ice Ic and Ih, respectively.
This error is in line with other state of the art machine learning interatomic potentials\cite{Niu20,Deringer20}.

\subsection{Enthalpy of fusion and heat capacities}

\begin{figure}[t]
\centering
  \includegraphics[width=0.55\textwidth]{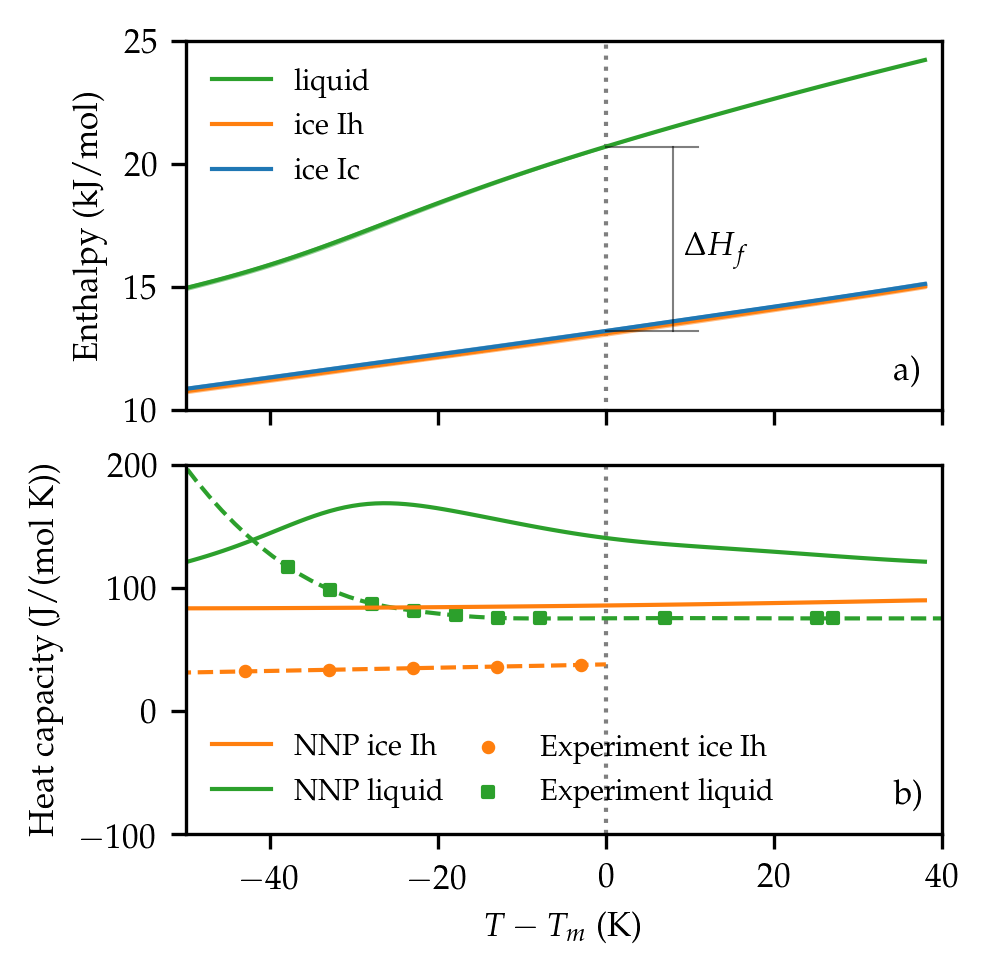}
  \caption{a) Enthalpy of liquid water, ice Ih, and ice Ic as a function of the temperature with respect to the melting temperature of ice Ih ($T_m$). The enthalpy of fusion $\Delta H_f$ is shown schematically. Errors were calculated from four-fold block averages and are shown as shaded regions (some are too small to be seen properly). b) Heat capacity of liquid water and ice Ih as a function of the temperature with respect to the melting temperature of ice Ih ($T_m$). Experimental results are shown with squares and circles, and the dashed lines correspond to splines fit to these data. Continuous lines are the results of calculation using the NNP model and were obtained from the enthalpies shown in a). Experimental data obtained from refs.\ \citenum{Chase86,Giauque36,Angell73}. Note that different melting temperatures were used for the NNP and the experimental results.}
  \label{fig:enthalpy_heat_capacity}
\end{figure}

We  calculated the enthalpy of liquid water, ice Ih, and ice Ic at atmospheric pressure from multithermal simulations of the NNP model in the interval 260-350 K using 64, 288, and 216 molecules respectively.
Multithermal simulations provide continuous data as a function of temperature and for this reason most results will be plotted as lines.
We employed four different proton configurations for the calculation of the enthalpy of the solid phases.
The enthalpies as a function of temperature are shown in Figure \ref{fig:enthalpy_heat_capacity}a.
In the next paragraphs we discuss the use of the computed enthalpies to calculate a variety of properties.

We first consider the enthalpy of fusion $\Delta H_f$, i.e. the enthalpy difference between liquid water and ice Ih at the melting temperature.
$\Delta H_f$ is shown schematically in Figure \ref{fig:enthalpy_heat_capacity}a.
Using a melting temperature of 312 K we obtain $\Delta H_f = 7.6$ kJ/mol which is $\sim$ 25\% higher than the experimental value $\Delta H_f = 6.01$ kJ/mol. 
We note that in $k_B T$ units $\Delta H_f$ is only $\sim$ 10 \% higher than the experimental value due to the higher melting temperature of the NNP.
A similar difference is found between the experimental $\Delta H_f$  and that of TIP4P/Ice.
We also calculated $\Delta H_f$ for SCAN DFT using Eq.~\eqref{eq:rew_dft} and found it to be similar to that of the NNP.
In Table \ref{tbl:enthalpy_fusion} we summarize the results of the enthalpy of fusion.

\begin{table}[b]
\small
  \caption{Enthalpy of fusion of ice Ih.}
  \label{tbl:enthalpy_fusion}
  \centering
  \begin{tabular*}{0.48\textwidth}{@{\extracolsep{\fill}}lll}
    \hline
     & $\Delta H_f$ (kJ/mol) & $\Delta H_f$ ($k_B T$) \\
    \hline
    NNP & 7.6(1) & 2.93(4) \\
    SCAN DFT & 7.5(4) & 2.9(2) \\
    TIP4P/Ice\cite{Abascal05} & 5.40 & 2.41 \\
    Experiment & 6.01 & 2.65 \\
    \hline
  \end{tabular*}
\end{table}

The enthalpy as a function of temperature can be used to calculate the heat capacity $C_p$ employing the relation $C_p = \partial H/ \partial T |_P $.
We plot $C_p$ as a function of temperature in Figure \ref{fig:enthalpy_heat_capacity}b and we compare the results with experimental data\cite{Chase86,Giauque36,Angell73}.
The heat capacity of ice Ic is not shown because it cannot be distinguished from that of ice Ih on the scale of the plot.
The heat capacity of ice Ih is somewhat higher than the experimental counterpart at all temperatures and it converges to the classical limit ($c_p=9 k_B$) towards low temperatures.
Nuclear quantum effects have been shown to decrease the heat capacity and bring it into closer agreement with experiment\cite{Vega10}.
The heat capacity of liquid water is also somewhat higher than in the experiment and exhibits a sharp increase as the temperature is lowered below the melting temperature.
This behavior mimics the experimental result although this increase appears at a lower temperature in the experiment.
There is also a maximum in the heat capacity at ambient pressure at around 280 K.
This maximum has been observed in other water models\cite{Gallo16} and hypothesized but not measured to date for real water.
The temperature shift observed here for the location of the heat capacity maximum has also been found for other properties of liquid water using the same NNP based on SCAN\cite{Gartner20}.
Also for liquid water nuclear quantum effects are likely to improve the agreement with experiment\cite{Vega10}.

\subsection{Densities}

\begin{figure}[t]
\centering
  \includegraphics[width=0.55\textwidth]{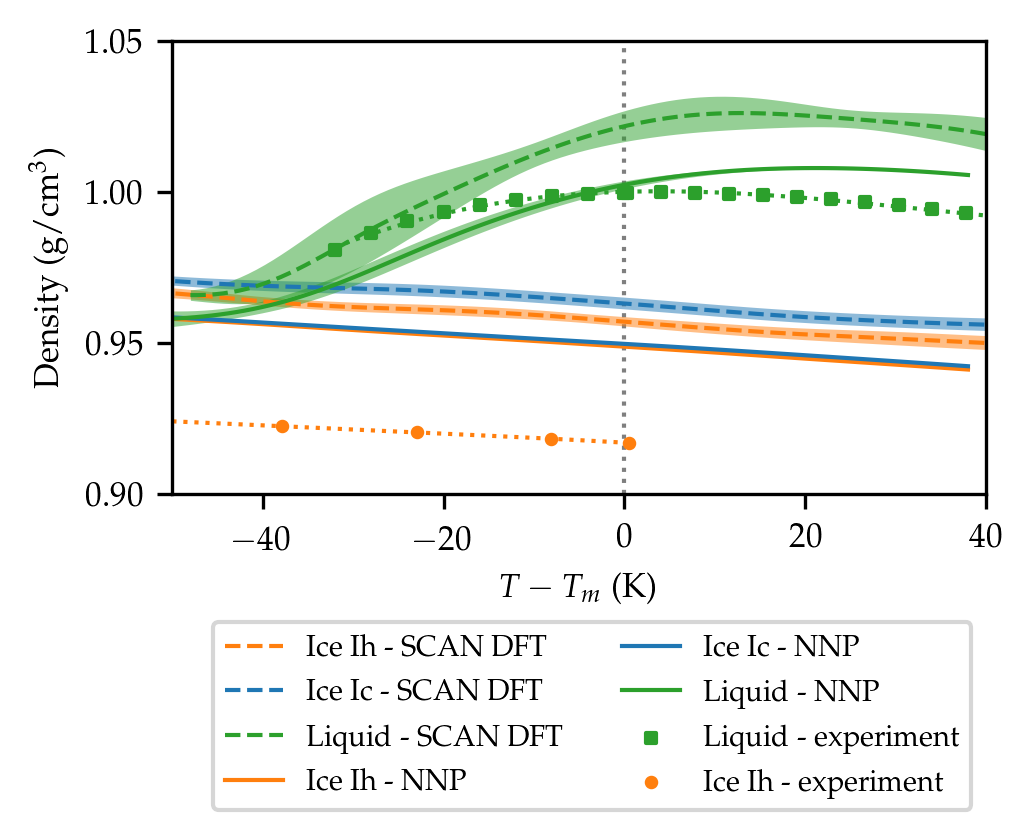}  \caption{Density of liquid water, ice Ih, and ice Ic as a function of the temperature with respect to the melting temperature of ice Ih ($T_m$). Experimental data obtained from refs.\ \citenum{Chase86,Hare87,Rottger94}. Note that different melting temperatures were used for the NNP, SCAN DFT, and the experimental results.}
  \label{fig:densities}
\end{figure}

From the multithermal simulations described above we can also calculate properties connected to the density of liquid water and ice.
In Figure \ref{fig:densities} we show the density of liquid water and ice Ih as a function of $T-T_m$, and compare them to experimental results\cite{Chase86,Hare87,Rottger94}.
The trends of the density as a function of temperature are correct for both phases.
The NNP correctly predicts that ice floats on water as was already noted in ref.~\citenum{Chen17}.
However, the change in density at melting is 9 \% for real water, and 6 \% in the NNP.
Densities of SCAN DFT water and ice Ih calculated using Eq.~\eqref{eq:rew_dft} differ somewhat from those of the NNP  and are also shown in Figure \ref{fig:densities}.
The agreement between the densities of the NNP and SCAN DFT would likely be improved by including the virial in the training procedure of the NNP.
The densities at the melting temperature are summarized in Table \ref{tbl:densities} and corresponding values for real water and the TIP4P/Ice are also provided.
\begin{table}[b]
\small
  \caption{Density of liquid water and ice Ih. Values reported at the melting temperature unless specified otherwise.}
  \label{tbl:densities}
  \centering
  \begin{tabular*}{0.48\textwidth}{@{\extracolsep{\fill}}ll}
    \hline
    \multicolumn{2}{c}{Ice Ih} \\
    \hline
     & $\rho_{Ih}$ (g/cm$^3$)  \\
    \hline
    NNP &  0.949(1) \\
    SCAN DFT (270 K) \cite{Chen17} & 0.96(1) \\
    SCAN DFT & 0.957(4) \\
    TIP4P/Ice\cite{Abascal05} &  0.906 \\
    Experiment &  0.917 \\
    \hline
    \multicolumn{2}{c}{Liquid water} \\
    \hline
     & $\rho_{l}$ (g/cm$^3$)  \\
    \hline
    NNP & 1.002(3) \\
    SCAN DFT (330 K) \cite{Chen17} & 1.05(3) \\
    SCAN DFT  & 1.020(5) \\
    TIP4P/Ice\cite{Abascal05}  &  0.985 \\
    Experiment & 0.999 \\
    \hline
    \multicolumn{2}{c}{Density change upon melting} \\
    \hline
     & $(\rho_{l}-\rho_{Ih})/\rho_{Ih}$ (\%)  \\
    \hline
    NNP &  6 \\
    SCAN DFT  &  6 \\
    TIP4P/Ice  &  9 \\
    Experiment &  9 \\
    \hline
  \end{tabular*}
\end{table}

The NNP also correctly predicts the existence of a density maximum as a function of temperature in liquid water.
This was already noted in ref.~\citenum{Gartner20}.
The NNP shows the density maximum at 333 K, 21 K above the melting temperature, while real water exhibits this maximum at around 277 K, only 4 K above the melting temperature.
We also used reweighting to obtain SCAN DFT results using Eq.~\eqref{eq:rew_dft}.
We obtained a temperature of maximum density of 321 K, 15 K above the melting temperature.
However, this property is affected by quantum fluctuations, as shown by the isotope effect.
We consider the difference between the temperature of maximum density of the liquid and the melting temperature of ice Ih $\Delta T = T_\mathrm{TMD} -T_m$.
$\Delta T$ is around 4 K for water but it has also been measured for heavy and tritiated water and $\Delta T$s of 7.4 K and 8.9 K were found, respectively\cite{NISTWebBook01}. 
Thus, for tritiated water, which is arguably the more classical-like isotopologue,  $\Delta T$ is not too far from the classical result of 15 K obtained here for SCAN.
These results are summarized in Table \ref{tbl:tmd}.

\begin{table}[b]
\small
  \caption{Temperature of maximum density of the liquid $T_\mathrm{TMD}$ and difference between the temperature of maximum density of the liquid and the melting temperature of ice Ih $\Delta T = T_\mathrm{TMD} -T_m$. We include the properties of heavy \ce{D2O} and tritiated \ce{T2O} water to illustrate that $\Delta T$ is larger the more classical the system. Experimental data from ref.~\citenum{NISTWebBook01}. }
  \label{tbl:tmd}
  \centering
  \begin{tabular*}{0.48\textwidth}{@{\extracolsep{\fill}}lll}
    \hline
     & $T_\mathrm{TMD}$ (K) & $\Delta T$ (K)  \\
    \hline
    NNP & 333(2) & 21(3) \\
    SCAN DFT & $\sim$ 321 & $\sim$ 15 \\
    TIP4P/Ice\cite{Abascal05}  & 295  & 23 \\
    Experiment \ce{H2O} & 277.1 & 4.0 \\
    Experiment \ce{D2O} & 284.3 & 7.4 \\    
    Experiment \ce{T2O} & 286.6 & 8.9 \\    

    \hline
  \end{tabular*}
\end{table}

\subsection{Supersaturation}

\begin{figure}[t]
\centering
  \includegraphics[width=0.55\textwidth]{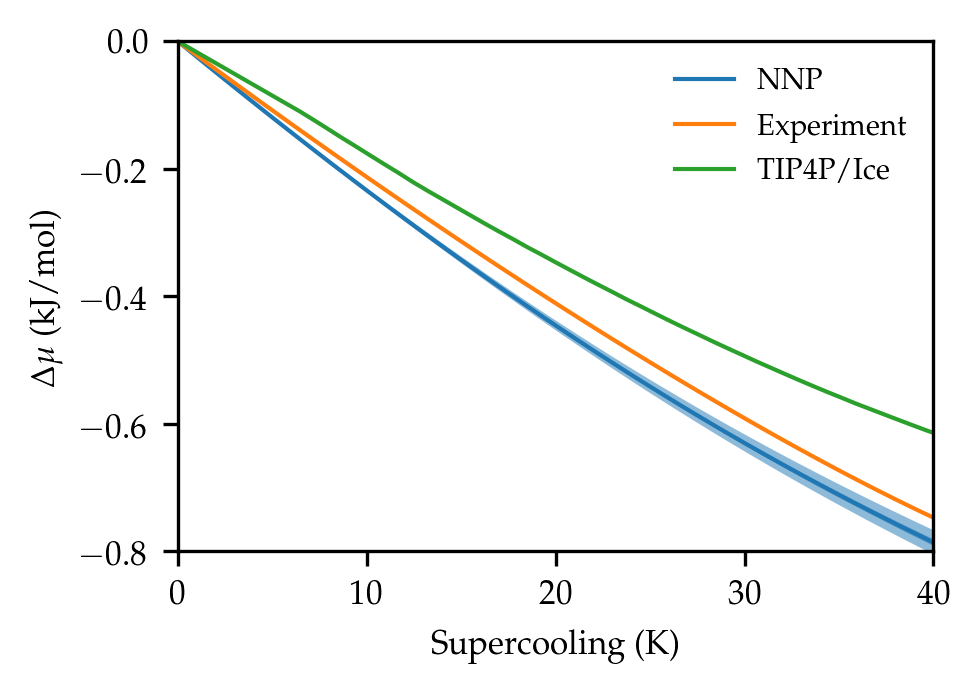}
  \caption{Driving force for nucleation. Calculations made with the NNP are compared against experimental results obtained from heat capacities reported in refs.\ \citenum{Chase86,Giauque36,Angell73} and calculations made with the semiempirical TIP4P/Ice model reported in ref.\ \citenum{Espinosa14}. The shaded blue area represents the error arising from the uncertainty in the enthalpies and in the melting temperature of ice Ih.}
  \label{fig:driving-force}
\end{figure}

We now set out to calculate the supersaturation $\Delta\mu_{l \rightarrow Ih}$, i.e.~the difference in chemical potential between ice Ih and liquid water, at greater undercoolings than those shown in Figure \ref{fig:free_energy}c. 
The supersaturation plays a central role in classical nucleation theory\cite{KashchievBook}, and affects the nucleation barrier, the nucleus size, and the nucleation rate.
Therefore, a precise estimation of $\Delta\mu_{l \rightarrow Ih}$ is needed to understand ice nucleation.
In order to calculate $\Delta\mu_{l \rightarrow Ih}$ we use the calculated enthalpies of liquid water $H_l$ and ice Ih $H_{Ih}$ and the thermodynamic relation,
\begin{equation}
\Delta\mu_{l \rightarrow Ih} (T) = T \int\limits_{T}^{T_m^{Ih}} \frac{H_{Ih}(T')-H_l(T')}{T'^2}dT'
\label{eq:chemical_potential_from_enthalpy}
\end{equation}
where $T_m^{Ih}$ is the melting temperature of ice Ih.
The results of these calculations are shown in Figure \ref{fig:driving-force}.
We also included the results for TIP4P/Ice from ref.\ \citenum{Espinosa14} and the experimental results calculated from the heat capacities shown in Figure \ref{fig:enthalpy_heat_capacity}b.
Our results show that at a 40 K undercooling the NNP overestimates the supersaturation by 5 \% while TIP4P/Ice underestimates it by around 18 \%.
The better agreement of the NNP with experiments is preserved even if $\Delta\mu_{l \rightarrow Ih}$ is considered in $k_B T$ units in which the different melting temperature might have an effect.
We also calculated $\Delta\mu_{l \rightarrow Ih}$ for SCAN DFT and found it to be equal to that of the NNP within the error bars of our calculation.

\subsection{Relative stability of ice Ih and ice Ic}

Another crucial characteristic of water that a model should capture is the relative stability of ice Ih and ice Ic.
Even though ice Ih is more stable than ice Ic at ambient pressure, there is significant evidence from simulations\cite{Haji15,Lupi17,Grabowska19,Niu19} and experiments\cite{Amaya17,Murray05} that nucleating ice clusters are rich in ice Ic at large supersaturations.
Furthermore, Quigley \cite{Quigley14} used a simple model to show that the fraction of cubic ice found in nucleating ice clusters depends significantly on the chemical potential difference between ice Ih and ice Ic.
This shows that a model that captures the relative stability between ice Ih and Ic is essential to understand the mechanism of ice nucleation from computer simulations.

\begin{figure}[t]
\centering
  \includegraphics[width=0.55\textwidth]{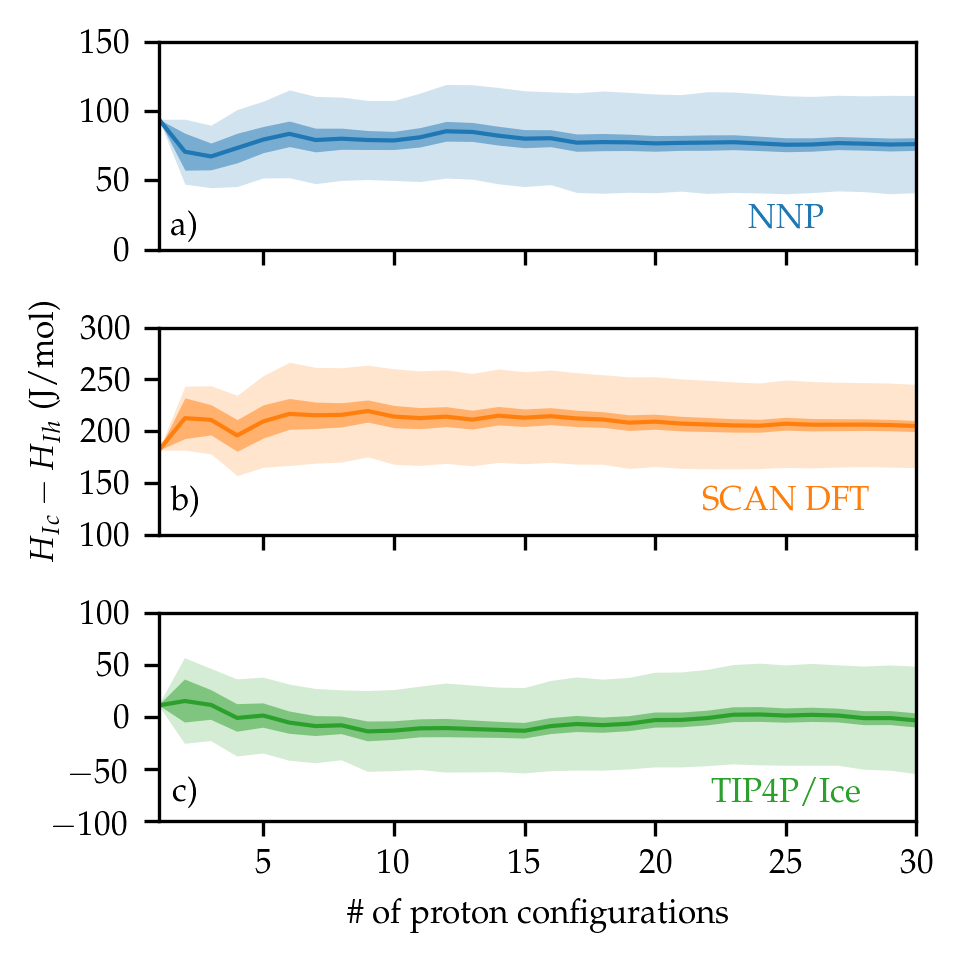}
  \caption{Enthalpy difference between cubic ($H_{Ic}$) and hexagonal ($H_{Ih}$) ice at 0 K as a function of the number of proton configurations considered for each polymorph. The line represents the mean value of $H_{Ic}-H_{Ih}$, the dark shaded region is the deviation of the mean, and the light shaded region is the deviation of the population. a) Enthalpies calculated with the NNP and systems of 432 and 512 molecules for ice Ih and Ic, respectively. b) Enthalpies calculated using SCAN DFT and systems of 128 and 64 molecules for ice Ih and Ic, respectively. c) Enthalpies calculated with TIP4P/Ice and systems of 432 and 512 molecules for ice Ih and Ic, respectively.}
  \label{fig:conv-proton-conf}
\end{figure}

To assess the stability of the polymorphs, we first calculated the enthalpies at 0 K using the NNP, and systems with 432 and 512 molecules for ice Ih and Ic, respectively.
The enthalpy can vary substantially for different proton configurations and for this reason we analyzed our results as a function of the number of proton configurations considered.
The difference in enthalpy between ice Ih and ice Ic, $H_{Ic}-H_{Ih}$, as a function of the number of proton configurations is shown in Fig.\ \ref{fig:conv-proton-conf}a.
It can be seen that around 10 proton configurations are needed for each polymorph in order to obtain a converged result.
From this calculation we obtain a mean of $H_{Ic}-H_{Ih}$ equal to 76 J/mol.
Results might differ significantly if only one proton configuration is considered.
The standard deviation of $H_{Ic}-H_{Ih}$ characterizes the variation of this quantity in different proton configurations and is also shown in Fig.\ \ref{fig:conv-proton-conf}.
We repeated the calculation for systems of 64 molecules and 128 molecules for ice Ic and Ih, respectively, and obtained identical results within the statistical error.
At variance with other properties, the calculation of $H_{Ic}-H_{Ih}$ for systems of 64 and 128 molecules is relatively inexpensive and can also be performed directly using DFT with the SCAN XC functional.
The results are reported in Fig.\ \ref{fig:conv-proton-conf}b.
The mean of $H_{Ic}-H_{Ih}$ within SCAN DFT is 205 J/mol.
We note that in this calculation we have ignored the contribution of the quantum mechanical zero point energy.
Furthermore, we found that $H_{Ic}-H_{Ih}$ within SCAN DFT depends somewhat on the plane wave kinetic energy cutoff.
We discuss this in detail in the ESI.
The NNP and SCAN DFT correctly predict that ice Ih is more stable than ice Ic at 0 K.
The results are also in good agreement with experimental calorimetric measurements that find values of $H_{Ic}-H_{Ih}$ around 50 J/mol\cite{Handa86,Mayer87,Yamamuro87}.
We note however that preparing samples of ice Ic is challenging experimentally and only recently it has become possible to obtain samples with high structural purity\cite{delRosso20,Salzmann20}.
It is also interesting to compare the NNP and SCAN DFT with other water models.
Thus, we calculated $H_{Ic}-H_{Ih}$ using TIP4P/Ice (shown in Fig.\ \ref{fig:conv-proton-conf}c) and obtained a result equal to zero within the error bars.
We also found that TIP4P/Ice results are somewhat sensitive to the accuracy in the determination of the long range electrostatic energy.
The value reported here corresponds to a typical accuracy used in molecular dynamics simulations.
We describe this and other details of the calculations with TIP4P/Ice in the ESI.
The results are summarized in Table \ref{tbl:enthalpy_diff_cub_hex}.

\begin{table}[b]
\small
  \caption{Relative stability of ice Ih and Ic. The temperature of the measurements or calculations is shown inside parenthesis.}
  \label{tbl:enthalpy_diff_cub_hex}
  \centering
  \begin{tabular*}{0.48\textwidth}{@{\extracolsep{\fill}}ll}
    \hline
     & $H_{Ic}-H_{Ih}$ (J/mol)\\
    \hline
    NNP  (0 K)  & 76(6) \\
    NNP (avg.\ 100-350 K)  & 103(21) \\
    SCAN DFT  (0 K) & 205(7) \\
    SCAN DFT  (avg.\ 100-350 K) & $\sim$83 \\
    Experiment\cite{Handa86} (200 K) & 51(2) \\
    Experiment\cite{Mayer87} (223 K) & 56   \\
    Experiment\cite{Yamamuro87} (200 K)& 37(1)   \\
    TIP4P/Ice (0 K) & 0(5) \\
    \hline
    & $\mu_{Ic}-\mu_{Ih}$ (J/mol)\\
    \hline
    NNP (312 K) & 65(37) \\
    NNP (avg.\ 100-350 K) & $\sim$87 \\
    SCAN DFT (avg.\ 100-350 K) & $\sim$117 \\
    Experiment\cite{Shilling06} (186 K) & 155(5)   \\
    TIP4P/Ice\cite{Zaragoza15} (200 K) & 0(20) \\
    \hline
  \end{tabular*}
\end{table}

\begin{figure}[t]
\centering
      \includegraphics[width=0.55\textwidth]{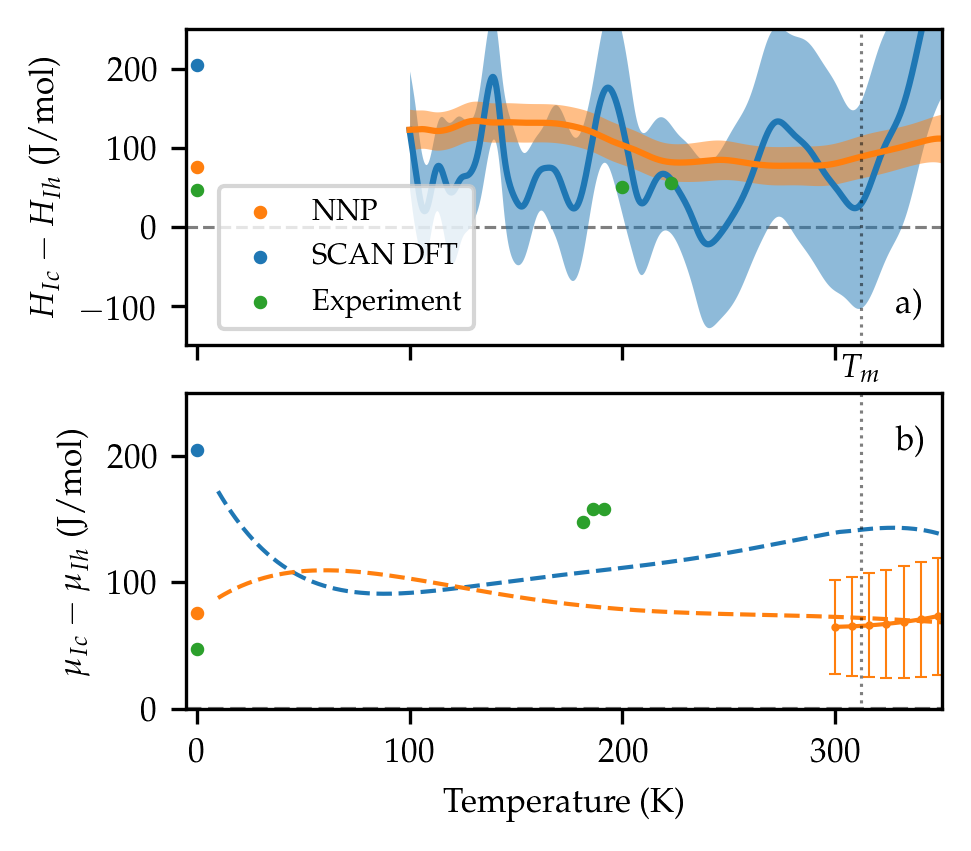}
  \caption{Relative stability of ice Ih and Ic at finite temperature. a) Difference in enthalpy between ice Ih and Ic  $H_{Ic}-H_{Ih}$. The results of the calculations in the interval 100-350 K are shown with lines. Error bars contain the statistical uncertainty in the calculation of the enthalpy at a fixed proton configuration and the variation due to different proton configurations (see the Appendix for details). Experimental results are shown in green circles and were taken from refs.\ \citenum{Handa86} and \citenum{Mayer87}. Zero temperature results are also shown in circles.   
  b) Difference in chemical potential between the polymorphs  $\mu_{Ic} - \mu_{Ih}$. The circles with errorbars are the results of the free energy calculations shown in Figure \ref{fig:free_energy}. The dashed lines were calculated using the enthalpies and Eq.\ \eqref{eq:diff_pot_chem_ices} (see text for details). The lines represent the most likely value for $\mu_{Ic} - \mu_{Ih}$.  Experimental results from refs.\ \citenum{Handa86} and \citenum{Shilling06} are shown as green circles.
  }
  \label{fig:ices_enthalpy_chempot}
\end{figure}

We now turn to study the stability of ice Ih and Ic at finite temperature.
We calculated $\Delta H_{Ih \rightarrow Ic} = H_{Ic}-H_{Ih}$ in the temperature interval 100-350 K from a new set of multithermal simulations using 64 and 128 water molecules for ice Ic and Ih, respectively, and 12 proton configurations for each polymorph.
$\Delta H_{Ih \rightarrow Ic}$ as a function of temperature is shown in Figure \ref{fig:ices_enthalpy_chempot}a where we also included the value at 0 K.
The results of the NNP are in good agreement with the experimental findings\cite{Handa86,Mayer87} with a mere 25 J/mol shift in enthalpy.
The enthalpy difference of SCAN DFT was obtained using reweighting and the corresponding results shown in Figure \ref{fig:ices_enthalpy_chempot}a are significantly more noisy than those of the NNP.
Nonetheless, the value of $\Delta H_{Ih \rightarrow Ic}$ for SCAN DFT averaged over all temperatures is 67 J/mol in very good agreement with the experiments.
We also analyzed the anharmonic effects in the enthalpy of ice Ih and ice Ic since Engel et al.\  found that anharmonic effects stabilize ice Ih with respect of ice Ic\cite{Engel15}.
This analysis is discussed in the ESI.
At the classical level studied here, we did not find evidence for a stabilization driven by anharmonic effects.
Classical SCAN thus predicts that ice Ih is more stable than ice Ic as a result of its higher stability at 0 K.

We then calculated the difference in chemical potential between the polymorphs  $\Delta \mu_{Ih \rightarrow Ic} = \mu_{Ic} - \mu_{Ih}$ using,
\begin{equation}
\Delta \mu_{Ih \rightarrow Ic}(T) = T \int\limits_{T}^{T_m^{Ic}} \frac{\Delta H_{l \rightarrow Ic}(T')}{T'^2}dT' 
  - T \int\limits_{T}^{T_m^{Ih}} \frac{\Delta H_{l \rightarrow Ih}(T')}{T'^2}dT' .
  \label{eq:diff_pot_chem_ices}
\end{equation}
Before performing the integration, we fit the enthalpy difference to a fifth-order polynomial in order to obtain a smooth $\Delta \mu_{Ih \rightarrow Ic}(T)$.
Furthermore, in the polynomial the linear term was set to zero to reflect that the classical heat capacities of both polymorphs are equal at 0 K.
The results are shown in Figure \ref{fig:ices_enthalpy_chempot}b both for the NNP and SCAN DFT.
The fact that $\Delta \mu_{Ih \rightarrow Ic}$ is larger for SCAN DFT than for the NNP at the ice Ih melting temperature is a result of the larger difference between the melting temperatures of ice Ih and Ic.
The values of $\Delta \mu_{Ih \rightarrow Ic}$ averaged over all studied temperatures are 87 J/mol for the NNP and 118 J/mol for SCAN DFT.
We also obtained an independent estimate of $\Delta \mu_{Ih \rightarrow Ic}$ from the crystallization simulations that is in good agreement with the results from Eq.~\eqref{eq:diff_pot_chem_ices} and is  shown in Figure \ref{fig:ices_enthalpy_chempot}b.
The results for the NNP and SCAN DFT are in very good agreement with the experiment.
However, the limited experimental data available suggests an enhancement of the stability of ice Ih as the temperature increases that was not observed here.
Furthermore, based on the results of refs.\ \citenum{Buxton19}, \citenum{Engel15} and \citenum{Cheng19} it would be possible that nuclear quantum effects could stabilize ice Ih further and bring the results of the NNP and SCAN DFT into even better agreement with experiment.
We stress that it is a remarkable success of SCAN that the sign and order of magnitude of $\Delta \mu_{Ih \rightarrow Ic}$ are in agreement with the experiment.
It is also interesting to compare this result with the stability of ice Ih and ice Ic in semiempirical potentials.
For instance, we recall that $\Delta \mu_{Ih \rightarrow Ic}$ is around 2 J/mol \cite{Quigley14,Cheng19} for the mW model, less than 20 J/mol for TIP4P/Ice\cite{Zaragoza15}, less than 6 J/mol for TIP4P/2005\cite{Zaragoza15}, and less than 10 J/mol \cite{Buxton19} for the flexible model TIP4P/F\cite{Habershon09}.

\section{Conclusions}

We have studied properties of water and ice relevant to ice nucleation as obtained with the SCAN exchange and correlation functional.
We found that the overall description of water and ice is excellent, and all experimental trends are captured qualitatively.
Furthermore, the performance of the semilocal SCAN density functional for the description of water and ice is similar to that of hybrid and van der Waals density functionals at a much reduced cost.

The melting temperature of ice Ih was obtained using enhanced sampling simulations that reproduce the crystallization process and validated using direct coexistence simulations.
Our estimate of the melting temperature of ice Ih in SCAN is 308 K, which is higher than the experimental value by 35 K.
This is a rather large difference but in line with many semiempirical water models\cite{Vega11}.
The NNP is a faithful representation of the SCAN DFT and the melting temperature is only 4 K higher.
The heat of fusion and the densities are also in very good agreement with the experiment.
One of the most remarkable successes of SCAN is the prediction of the supersaturation, i.e. difference in chemical potential between liquid water and ice Ih, within 5 \% of the correct value at relatively large supercoolings (40 K).
The agreement is better than in the TIP4P family that has an error of around 20 \%\cite{Espinosa14}.
Another striking success of SCAN is the prediction that ice Ih is more stable than ice Ic.
We found that the difference in enthalpy between ice Ic and Ih is in excellent agreement with values around 50 J/mol found in experiments.
One may argue that this is a rather small energy difference.
However, it is an order of magnitude larger than the difference in enthalpy between ice Ic and Ih in semiempirical potentials, such as mW and TIP4P/Ice.
Furthermore, this small but crucial energy difference gives rise to the snowflakes with six-fold symmetry found in nature.
The availability of DFT functionals and machine learning potentials able to describe subtle energy differences is not only relevant for ices but also in other systems that exhibit rich polymorphism such as molecular crystals.

We also recall other properties of SCAN water calculated elsewhere.
The diffusion coefficient of water is in good agreement with the experiment if the temperatures are shifted to take into account the difference in melting temperature between the experiment and SCAN\cite{Gartner20}.
This property is relevant for the kinetics of nucleation and thus important to the topic of this work.
It was also shown in ref.~\citenum{Chen17} that the structure of water represented by the radial distribution function is captured well by SCAN.
Finally, we note that SCAN has also been shown to reproduce some important anomalies of liquid water such as the density maximum, the sharp increase in heat capacity and isothermal compressibility, and the possible existence of a liquid-liquid critical point at deeply supercooled conditions\cite{Gartner20}.

We conclude with a summary of the strategy that we have used to calculate complex properties at the DFT level. This strategy is based on: 1) driving the dynamics using a machine learning interatomic potential as a surrogate for the DFT model and leveraging enhanced sampling, 2) extracting configurations from these simulations and calculating the energy at the DFT level, and 3) using a reweighting procedure to calculate ensemble averages of the DFT model.
This route can be more efficient than driving dynamics with forces calculated on-the-fly with DFT.
The availability of a machine learning potential that reproduces well the DFT energies is crucial to the success of this approach.
Here we have employed a NNP based on the DeePMD framework and we have shown that it provides a remarkably faithful representation of the SCAN potential energy surface.

\section*{Acknowledgement}

P.M.P thanks Zachary Goldsmith for assistance in setting up the DFT calculations, and Linfeng Zhang and Thomas Gartner for sharing their neural network potential for water.
P.M.P was supported by an Early Postdoc.Mobility fellowship from the Swiss National Science Foundation. 
This work was conducted within the center: Chemistry in Solution and at Interfaces funded by the DoE under Award DE-SC0019394.
This research used resources of the National Energy Research Scientific Computing Center (NERSC), a U.S. Department of Energy Office of Science User Facility operated under Contract No. DE-AC02-05CH11231.
Simulations reported here were substantially performed using the Princeton Research Computing resources at Princeton University which is consortium of groups including the Princeton Institute for Computational Science and Engineering and the Princeton University Office of Information Technology's Research Computing department.

\section*{Supplementary information}

Further computational details and miscellaneous analysis.

\section*{Data availability statement}
The input files and results of the simulations are openly available on GitHub\cite{GitHubRepo} and on PLUMED-NEST\cite{Bonomi19} (\url{www.plumed-nest.org}), the public repository of the PLUMED consortium, as plumID:21.002.
The NNP is available at \url{https://doi.org/10.34770/45m3-am91}.

\section*{Appendix: Error from averaging over proton configurations }

Taking into account proton disorder is crucial to calculate properties of ice Ih and Ic.
We consider the calculation of a property $P$ that due to proton disorder can be modelled as a random variable.
We assume that $P$ is normally distributed with mean $\mu_P$ and standard deviation $\sigma_{P}$.
We calculate $P$ by performing $M$ simulations with different proton configurations.
Each simulation provides an estimate $p_i$ of the property $P$ for each proton configuration $i=1,...,M$ and a statistical error $\sigma_{p_i}$ in the estimation of $p_i$.
The estimator of $\mu_P$ is,
\begin{equation}
    \bar{p}=\frac{1}{M}\sum\limits_{i=1}^M p_i .
\end{equation}
However, the fact that $p_i$ are determined using sampling means that their value is also a random variable.
We will define $\delta p_i$ as the difference between the true $p_i$ and the ones estimated through sampling.
The estimator of $\mu_P$ then becomes,
\begin{equation}
    \bar{p}=\frac{1}{M}\sum\limits_{i=1}^M p_i +\delta p_i .
\end{equation}
We also consider that the error $\delta p_i$ is normally distributed with mean zero and standard deviation $\sigma_{p_i}$.
We now can use the fact that the standard deviation $\sigma$ of a sum of independent normally distributed random variables with standard deviations $\sigma_1$ and $\sigma_2$ is $\sigma=\sqrt{\sigma_1^2+\sigma_2^2}$.
Then,
\begin{equation}
    \sigma_{\bar{p}}=\sqrt{\frac{\sigma_{P}^2}{M} + \frac{ \sum_{i=1}^M \sigma_{p_i}^2}{M^2} }.
\end{equation}
This equation clearly shows one contribution to the error from the finite sampling of the proton configurations and another from the finite sampling of the configuration space for a fixed proton configuration.
$\sigma_{P}$ can be calculated using the estimator of the standard deviation of the $p_i$, i.e:
\begin{equation}
    \hat{\sigma}_P^2 = \frac{\sum_{i=1}^M (p_i-\mu_P)^2}{M-1} \approx \sigma_P^2
\end{equation}
and $\sigma_{p_i}$ can be obtained using block averages as outlined in appendix B of ref.~\citenum{Invernizzi20}.

\bibliographystyle{unsrt}

\begin{thebibliography}{10}

\bibitem{Debenedetti03}
Pablo~G Debenedetti.
\newblock Supercooled and glassy water.
\newblock {\em Journal of Physics: Condensed Matter}, 15(45):R1669, 2003.

\bibitem{Matsumoto02}
Masakazu Matsumoto, Shinji Saito, and Iwao Ohmine.
\newblock Molecular dynamics simulation of the ice nucleation and growth
  process leading to water freezing.
\newblock {\em Nature}, 416(6879):409--413, 2002.

\bibitem{Molinero09}
Valeria Molinero and Emily~B Moore.
\newblock Water modeled as an intermediate element between carbon and silicon.
\newblock {\em The Journal of Physical Chemistry B}, 113(13):4008--4016, 2009.

\bibitem{Abascal05}
JLF Abascal, E~Sanz, R~Garc{\'\i}a~Fern{\'a}ndez, and C~Vega.
\newblock A potential model for the study of ices and amorphous water:
  Tip4p/ice.
\newblock {\em The Journal of chemical physics}, 122(23):234511, 2005.

\bibitem{Sanz13}
Eduardo Sanz, Carlos Vega, JR~Espinosa, R~Caballero-Bernal, JLF Abascal, and
  C~Valeriani.
\newblock Homogeneous ice nucleation at moderate supercooling from molecular
  simulation.
\newblock {\em Journal of the American Chemical Society}, 135(40):15008--15017,
  2013.

\bibitem{Espinosa14}
JR~Espinosa, E~Sanz, C~Valeriani, and C~Vega.
\newblock Homogeneous ice nucleation evaluated for several water models.
\newblock {\em The Journal of chemical physics}, 141(18):18C529, 2014.

\bibitem{Li11}
Tianshu Li, Davide Donadio, Giovanna Russo, and Giulia Galli.
\newblock Homogeneous ice nucleation from supercooled water.
\newblock {\em Physical Chemistry Chemical Physics}, 13(44):19807--19813, 2011.

\bibitem{Yan12}
JY~Yan and GN~Patey.
\newblock Molecular dynamics simulations of ice nucleation by electric fields.
\newblock {\em The Journal of Physical Chemistry A}, 116(26):7057--7064, 2012.

\bibitem{Bauerecker08}
Sigurd Bauerecker, Peter Ulbig, Victoria Buch, Lubo{\v{s}} Vrbka, and Pavel
  Jungwirth.
\newblock Monitoring ice nucleation in pure and salty water via high-speed
  imaging and computer simulations.
\newblock {\em The Journal of Physical Chemistry C}, 112(20):7631--7636, 2008.

\bibitem{Quigley08}
D~Quigley and PM~Rodger.
\newblock Metadynamics simulations of ice nucleation and growth.
\newblock {\em The Journal of chemical physics}, 128(15):154518, 2008.

\bibitem{Debenedetti20}
Pablo~G Debenedetti, Francesco Sciortino, and G{\"u}l~H Zerze.
\newblock Second critical point in two realistic models of water.
\newblock {\em Science}, 369(6501):289--292, 2020.

\bibitem{Vega11}
Carlos Vega and Jose~LF Abascal.
\newblock Simulating water with rigid non-polarizable models: a general
  perspective.
\newblock {\em Physical Chemistry Chemical Physics}, 13(44):19663--19688, 2011.

\bibitem{Kohn65}
Walter Kohn and Lu~Jeu Sham.
\newblock Self-consistent equations including exchange and correlation effects.
\newblock {\em Physical review}, 140(4A):A1133, 1965.

\bibitem{KohanoffBook}
Jorge Kohanoff.
\newblock {\em Electronic structure calculations for solids and molecules:
  theory and computational methods}.
\newblock Cambridge University Press, 2006.

\bibitem{Gillan16}
Michael~J Gillan, Dario Alf{\`e}, and Angelos Michaelides.
\newblock Perspective: How good is dft for water?
\newblock {\em The Journal of chemical physics}, 144(13):130901, 2016.

\bibitem{Perdew96}
John~P Perdew, Kieron Burke, and Matthias Ernzerhof.
\newblock Generalized gradient approximation made simple.
\newblock {\em Physical review letters}, 77(18):3865, 1996.

\bibitem{Schmidt09}
Jochen Schmidt, Joost VandeVondele, I-F~William Kuo, Daniel Sebastiani, J~Ilja
  Siepmann, J{\"u}rg Hutter, and Christopher~J Mundy.
\newblock Isobaric- isothermal molecular dynamics simulations utilizing density
  functional theory: an assessment of the structure and density of water at
  near-ambient conditions.
\newblock {\em The Journal of Physical Chemistry B}, 113(35):11959--11964,
  2009.

\bibitem{Gaiduk15}
Alex~P Gaiduk, Francois Gygi, and Giulia Galli.
\newblock Density and compressibility of liquid water and ice from
  first-principles simulations with hybrid functionals.
\newblock {\em The journal of physical chemistry letters}, 6(15):2902--2908,
  2015.

\bibitem{Chen17}
Mohan Chen, Hsin-Yu Ko, Richard~C Remsing, Marcos F~Calegari Andrade, Biswajit
  Santra, Zhaoru Sun, Annabella Selloni, Roberto Car, Michael~L Klein, John~P
  Perdew, et~al.
\newblock Ab initio theory and modeling of water.
\newblock {\em Proceedings of the National Academy of Sciences},
  114(41):10846--10851, 2017.

\bibitem{Lin12}
I-Chun Lin, Ari~P Seitsonen, Ivano Tavernelli, and Ursula Rothlisberger.
\newblock Structure and dynamics of liquid water from ab initio molecular
  dynamics - comparison of blyp, pbe, and revpbe density functionals with and
  without van der waals corrections.
\newblock {\em Journal of chemical theory and computation}, 8(10):3902--3910,
  2012.

\bibitem{Cheng19}
Bingqing Cheng, Edgar~A Engel, J{\"o}rg Behler, Christoph Dellago, and Michele
  Ceriotti.
\newblock Ab initio thermodynamics of liquid and solid water.
\newblock {\em Proceedings of the National Academy of Sciences},
  116(4):1110--1115, 2019.

\bibitem{Sun15}
Jianwei Sun, Adrienn Ruzsinszky, and John~P Perdew.
\newblock Strongly constrained and appropriately normed semilocal density
  functional.
\newblock {\em Physical review letters}, 115(3):036402, 2015.

\bibitem{Sun16}
Jianwei Sun, Richard~C Remsing, Yubo Zhang, Zhaoru Sun, Adrienn Ruzsinszky,
  Haowei Peng, Zenghui Yang, Arpita Paul, Umesh Waghmare, Xifan Wu, et~al.
\newblock Accurate first-principles structures and energies of diversely bonded
  systems from an efficient density functional.
\newblock {\em Nature chemistry}, 8(9):831, 2016.

\bibitem{Car85}
Richard Car and Mark Parrinello.
\newblock Unified approach for molecular dynamics and density-functional
  theory.
\newblock {\em Physical review letters}, 55(22):2471, 1985.

\bibitem{Soper08}
AK~Soper and CJ~Benmore.
\newblock Quantum differences between heavy and light water.
\newblock {\em Physical review letters}, 101(6):065502, 2008.

\bibitem{Yao20}
Yi~Yao and Yosuke Kanai.
\newblock Temperature dependence of nuclear quantum effects on liquid water via
  artificial neural network model based on scan meta-gga functional.
\newblock {\em The Journal of Chemical Physics}, 153(4):044114, 2020.

\bibitem{Gartner20}
Thomas~E Gartner, Linfeng Zhang, Pablo~M Piaggi, Roberto Car, Athanassios~Z
  Panagiotopoulos, and Pablo~G Debenedetti.
\newblock Signatures of a liquid--liquid transition in an ab initio deep neural
  network model for water.
\newblock {\em Proceedings of the National Academy of Sciences},
  117(42):26040--26046, 2020.

\bibitem{Plimpton95}
Steve Plimpton.
\newblock Fast parallel algorithms for short-range molecular dynamics.
\newblock {\em Journal of computational physics}, 117(1):1--19, 1995.

\bibitem{Wang18}
Han Wang, Linfeng Zhang, Jiequn Han, and E~Weinan.
\newblock Deepmd-kit: A deep learning package for many-body potential energy
  representation and molecular dynamics.
\newblock {\em Computer Physics Communications}, 228:178--184, 2018.

\bibitem{Zhang18}
Linfeng Zhang, Jiequn Han, Han Wang, Roberto Car, and E~Weinan.
\newblock Deep potential molecular dynamics: a scalable model with the accuracy
  of quantum mechanics.
\newblock {\em Physical review letters}, 120(14):143001, 2018.

\bibitem{Zhang18end}
Linfeng Zhang, Jiequn Han, Han Wang, Wissam Saidi, Roberto Car, et~al.
\newblock End-to-end symmetry preserving inter-atomic potential energy model
  for finite and extended systems.
\newblock {\em Advances in Neural Information Processing Systems},
  31:4436--4446, 2018.

\bibitem{Zhang20}
Yuzhi Zhang, Haidi Wang, Weijie Chen, Jinzhe Zeng, Linfeng Zhang, Han Wang, and
  E~Weinan.
\newblock Dp-gen: A concurrent learning platform for the generation of reliable
  deep learning based potential energy models.
\newblock {\em Computer Physics Communications}, page 107206, 2020.

\bibitem{Zhang19}
Linfeng Zhang, De-Ye Lin, Han Wang, Roberto Car, and E~Weinan.
\newblock Active learning of uniformly accurate interatomic potentials for
  materials simulation.
\newblock {\em Physical Review Materials}, 3(2):023804, 2019.

\bibitem{Lu20}
Denghui Lu, Han Wang, Mohan Chen, Jiduan Liu, Lin Lin, Roberto Car, E~Weinan,
  Weile Jia, and Linfeng Zhang.
\newblock 86 pflops deep potential molecular dynamics simulation of 100 million
  atoms with ab initio accuracy.
\newblock {\em Computer Physics Communications}, page 107624, 2020.

\bibitem{Bussi07}
Giovanni Bussi, Davide Donadio, and Michele Parrinello.
\newblock Canonical sampling through velocity rescaling.
\newblock {\em The Journal of chemical physics}, 126(1):014101, 2007.

\bibitem{Parrinello81}
Michele Parrinello and Aneesur Rahman.
\newblock Polymorphic transitions in single crystals: A new molecular dynamics
  method.
\newblock {\em Journal of Applied physics}, 52(12):7182--7190, 1981.

\bibitem{Matsumoto18}
Masakazu Matsumoto, Takuma Yagasaki, and Hideki Tanaka.
\newblock Genice: Hydrogen-disordered ice generator.
\newblock {\em Journal of computational chemistry}, 39(1):61--64, 2018.

\bibitem{Conde17}
MM~Conde, M~Rovere, and P~Gallo.
\newblock High precision determination of the melting points of water
  tip4p/2005 and water tip4p/ice models by the direct coexistence technique.
\newblock {\em The Journal of chemical physics}, 147(24):244506, 2017.

\bibitem{Laio02}
Alessandro Laio and Michele Parrinello.
\newblock Escaping free-energy minima.
\newblock {\em Proceedings of the National Academy of Sciences},
  99(20):12562--12566, 2002.

\bibitem{Barducci08}
Alessandro Barducci, Giovanni Bussi, and Michele Parrinello.
\newblock Well-tempered metadynamics: A smoothly converging and tunable
  free-energy method.
\newblock {\em Physical review letters}, 100(2):020603, 2008.

\bibitem{Valsson14}
Omar Valsson and Michele Parrinello.
\newblock Variational approach to enhanced sampling and free energy
  calculations.
\newblock {\em Physical review letters}, 113(9):090601, 2014.

\bibitem{Invernizzi20rethinking}
Michele Invernizzi and Michele Parrinello.
\newblock Rethinking metadynamics: from bias potentials to probability
  distributions.
\newblock {\em The Journal of Physical Chemistry Letters}, 11(7):2731--2736,
  2020.

\bibitem{Tribello14}
Gareth~A Tribello, Massimiliano Bonomi, Davide Branduardi, Carlo Camilloni, and
  Giovanni Bussi.
\newblock Plumed 2: New feathers for an old bird.
\newblock {\em Computer Physics Communications}, 185(2):604--613, 2014.

\bibitem{Bonomi19}
Massimiliano Bonomi, Giovanni Bussi, Carlo Camilloni, Gareth~A Tribello, Pavel
  Ban{\'a}{\v{s}}, Alessandro Barducci, Mattia Bernetti, Peter~G Bolhuis,
  Sandro Bottaro, Davide Branduardi, et~al.
\newblock Promoting transparency and reproducibility in enhanced molecular
  simulations.
\newblock {\em Nature methods}, 16(8):670--673, 2019.

\bibitem{Piaggi19}
Pablo~M Piaggi and Michele Parrinello.
\newblock Multithermal-multibaric molecular simulations from a variational
  principle.
\newblock {\em Physical Review Letters}, 122(5):050601, 2019.

\bibitem{Invernizzi20}
Michele Invernizzi, Pablo~M. Piaggi, and Michele Parrinello.
\newblock Unified approach to enhanced sampling.
\newblock {\em Phys. Rev. X}, 10:041034, Nov 2020.

\bibitem{Piaggi20}
Pablo~M Piaggi and Roberto Car.
\newblock Phase equilibrium of liquid water and hexagonal ice from enhanced
  sampling molecular dynamics simulations.
\newblock {\em The Journal of Chemical Physics}, 152(20):204116, 2020.

\bibitem{Piaggi19b}
Pablo~M Piaggi and Michele Parrinello.
\newblock Calculation of phase diagrams in the multithermal-multibaric
  ensemble.
\newblock {\em The Journal of chemical physics}, 150(24):244119, 2019.

\bibitem{vescode}
\textit{VES Code}, a library that implements enhanced sampling methods based on
  Variationally Enhanced Sampling written by O.\ Valsson. For the current
  version, see http://www.ves-code.org.

\bibitem{Giannozzi09}
Paolo Giannozzi, Stefano Baroni, Nicola Bonini, Matteo Calandra, Roberto Car,
  Carlo Cavazzoni, Davide Ceresoli, Guido~L Chiarotti, Matteo Cococcioni,
  Ismaila Dabo, et~al.
\newblock Quantum espresso: a modular and open-source software project for
  quantum simulations of materials.
\newblock {\em Journal of physics: Condensed matter}, 21(39):395502, 2009.

\bibitem{Giannozzi17}
Paolo Giannozzi, Oliviero Andreussi, Thomas Brumme, Oana Bunau, M~Buongiorno
  Nardelli, Matteo Calandra, Roberto Car, Carlo Cavazzoni, Davide Ceresoli,
  Matteo Cococcioni, et~al.
\newblock Advanced capabilities for materials modelling with quantum espresso.
\newblock {\em Journal of Physics: Condensed Matter}, 29(46):465901, 2017.

\bibitem{Marques12}
Miguel~AL Marques, Micael~JT Oliveira, and Tobias Burnus.
\newblock Libxc: A library of exchange and correlation functionals for density
  functional theory.
\newblock {\em Computer physics communications}, 183(10):2272--2281, 2012.

\bibitem{Hamann13}
DR~Hamann.
\newblock Optimized norm-conserving vanderbilt pseudopotentials.
\newblock {\em Physical Review B}, 88(8):085117, 2013.

\bibitem{Zwanzig54}
Robert~W Zwanzig.
\newblock High-temperature equation of state by a perturbation method. i.
  nonpolar gases.
\newblock {\em The Journal of Chemical Physics}, 22(8):1420--1426, 1954.

\bibitem{Chehaibou19}
Bilal Chehaibou, Michael Badawi, Tomá\v{s} Bu\v{c}ko, Timur Bazhirov, and
  Dario Rocca.
\newblock Computing rpa adsorption enthalpies by machine learning thermodynamic
  perturbation theory.
\newblock {\em Journal of Chemical Theory and Computation}, 15(11):6333--6342,
  2019.

\bibitem{Binder87}
Kurt Binder.
\newblock Theory of first-order phase transitions.
\newblock {\em Reports on progress in physics}, 50(7):783, 1987.

\bibitem{Kong94}
Augustine Kong, Jun~S Liu, and Wing~Hung Wong.
\newblock Sequential imputations and bayesian missing data problems.
\newblock {\em Journal of the American statistical association},
  89(425):278--288, 1994.

\bibitem{Elvira18}
V{\'\i}ctor Elvira, Luca Martino, and Christian~P Robert.
\newblock Rethinking the effective sample size.
\newblock {\em arXiv preprint arXiv:1809.04129}, 2018.

\bibitem{Niu20}
Haiyang Niu, Luigi Bonati, Pablo~M Piaggi, and Michele Parrinello.
\newblock Ab initio phase diagram and nucleation of gallium.
\newblock {\em Nature Communications}, 11(1):1--9, 2020.

\bibitem{Deringer20}
Volker~L Deringer, Miguel~A Caro, and G{\'a}bor Cs{\'a}nyi.
\newblock A general-purpose machine-learning force field for bulk and
  nanostructured phosphorus.
\newblock {\em Nature communications}, 11(1):1--11, 2020.

\bibitem{Chase86}
Malcolm~W Chase.
\newblock Janaf thermochemical tables.
\newblock {\em JANAF thermochemical tables, by Chase, MW Washington, DC:
  American Chemical Society; New York: American Institute of Physics for the
  National Bureau of Standards, c1986.. United States. National Bureau of
  Standards.}, 1, 1986.

\bibitem{Giauque36}
WF~Giauque and JW~Stout.
\newblock The entropy of water and the third law of thermodynamics. the heat
  capacity of ice from 15 to 273 k.
\newblock {\em Journal of the American Chemical Society}, 58(7):1144--1150,
  1936.

\bibitem{Angell73}
CA~Angell, J~Shuppert, and JC~Tucker.
\newblock Anomalous properties of supercooled water. heat capacity,
  expansivity, and proton magnetic resonance chemical shift from 0 to-38\%.
\newblock {\em The Journal of Physical Chemistry}, 77(26):3092--3099, 1973.

\bibitem{Vega10}
C~Vega, Mar{\'\i}a~M Conde, Carl McBride, Jos{\'e} Luis~F Abascal, Eva~G Noya,
  Rafael Ram{\'\i}rez, and Luis~M Ses{\'e}.
\newblock Heat capacity of water: A signature of nuclear quantum effects.
\newblock {\em The Journal of chemical physics}, 132(4):046101, 2010.

\bibitem{Gallo16}
Paola Gallo, Katrin Amann-Winkel, Charles~Austen Angell, Mikhail~Alexeevich
  Anisimov, Fr{'e}d{'e}ric Caupin, Charusita Chakravarty, Erik Lascaris, Thomas
  Loerting, Athanassios~Zois Panagiotopoulos, John Russo, et~al.
\newblock Water: A tale of two liquids.
\newblock {\em Chemical reviews}, 116(13):7463--7500, 2016.

\bibitem{Hare87}
DE~Hare and CM~Sorensen.
\newblock The density of supercooled water. ii. bulk samples cooled to the
  homogeneous nucleation limit.
\newblock {\em The Journal of chemical physics}, 87(8):4840--4845, 1987.

\bibitem{Rottger94}
K~R{\"o}ttger, A~Endriss, J{\"o}rg Ihringer, S~Doyle, and WF~Kuhs.
\newblock Lattice constants and thermal expansion of h2o and d2o ice ih between
  10 and 265 k.
\newblock {\em Acta Crystallographica Section B: Structural Science},
  50(6):644--648, 1994.

\bibitem{NISTWebBook01}
Peter~J Linstrom and William~G Mallard.
\newblock The nist chemistry webbook: A chemical data resource on the internet.
\newblock {\em Journal of Chemical \& Engineering Data}, 46(5):1059--1063,
  2001.

\bibitem{KashchievBook}
Dimo Kashchiev.
\newblock {\em Nucleation: Basic Theory with Applications}.
\newblock Butterworth-Heinemann, Oxford, 2000.

\bibitem{Haji15}
Amir Haji-Akbari and Pablo~G Debenedetti.
\newblock Direct calculation of ice homogeneous nucleation rate for a molecular
  model of water.
\newblock {\em Proceedings of the National Academy of Sciences},
  112(34):10582--10588, 2015.

\bibitem{Lupi17}
Laura Lupi, Arpa Hudait, Baron Peters, Michael Gr{\"u}nwald, Ryan~Gotchy
  Mullen, Andrew~H Nguyen, and Valeria Molinero.
\newblock Role of stacking disorder in ice nucleation.
\newblock {\em Nature}, 551(7679):218--222, 2017.

\bibitem{Grabowska19}
Joanna Grabowska.
\newblock Why is the cubic structure preferred in newly formed ice?
\newblock {\em Physical Chemistry Chemical Physics}, 21(33):18043--18047, 2019.

\bibitem{Niu19}
Haiyang Niu, Yi~Isaac Yang, and Michele Parrinello.
\newblock Temperature dependence of homogeneous nucleation in ice.
\newblock {\em Physical review letters}, 122(24):245501, 2019.

\bibitem{Amaya17}
Andrew~J Amaya, Harshad Pathak, Viraj~P Modak, Hartawan Laksmono, N~Duane Loh,
  Jonas~A Sellberg, Raymond~G Sierra, Trevor~A McQueen, Matt~J Hayes, Garth~J
  Williams, et~al.
\newblock How cubic can ice be?
\newblock {\em The Journal of Physical Chemistry Letters}, 8(14):3216--3222,
  2017.

\bibitem{Murray05}
Benjamin~J Murray, Daniel~A Knopf, and Allan~K Bertram.
\newblock The formation of cubic ice under conditions relevant to earth's
  atmosphere.
\newblock {\em Nature}, 434(7030):202--205, 2005.

\bibitem{Quigley14}
D~Quigley.
\newblock Communication: Thermodynamics of stacking disorder in ice nuclei,
  2014.

\bibitem{Handa86}
Y~Paul Handa, DD~Klug, and Edward Whalley.
\newblock Difference in energy between cubic and hexagonal ice.
\newblock {\em The Journal of chemical physics}, 84(12):7009--7010, 1986.

\bibitem{Mayer87}
Erwin Mayer and Andreas Hallbrucker.
\newblock Cubic ice from liquid water.
\newblock {\em Nature}, 325(6105):601--602, 1987.

\bibitem{Yamamuro87}
O~Yamamuro, M~Oguni, T~Matsuo, and H~Suga.
\newblock Heat capacity and glass transition of pure and doped cubic ices.
\newblock {\em Journal of Physics and Chemistry of Solids}, 48(10):935--942,
  1987.

\bibitem{delRosso20}
Leonardo Del~Rosso, Milva Celli, Francesco Grazzi, Michele Catti, Thomas~C
  Hansen, A~Dominic Fortes, and Lorenzo Ulivi.
\newblock Cubic ice ic without stacking defects obtained from ice xvii.
\newblock {\em Nature Materials}, pages 1--6, 2020.

\bibitem{Salzmann20}
Christoph~G Salzmann and Benjamin~J Murray.
\newblock Ice goes fully cubic.
\newblock {\em Nature Materials}, 19(6):586--587, 2020.

\bibitem{Shilling06}
JE~Shilling, MA~Tolbert, OB~Toon, EJ~Jensen, Benjamin~J Murray, and Allan~K
  Bertram.
\newblock Measurements of the vapor pressure of cubic ice and their
  implications for atmospheric ice clouds.
\newblock {\em Geophysical research letters}, 33(17), 2006.

\bibitem{Zaragoza15}
Alberto Zaragoza, Maria~M Conde, Jorge~R Espinosa, Chantal Valeriani, Carlos
  Vega, and Eduardo Sanz.
\newblock Competition between ices ih and ic in homogeneous water freezing.
\newblock {\em The Journal of Chemical Physics}, 143(13):134504, 2015.

\bibitem{Engel15}
Edgar~A Engel, Bartomeu Monserrat, and Richard~J Needs.
\newblock Anharmonic nuclear motion and the relative stability of hexagonal and
  cubic ice.
\newblock {\em Physical Review X}, 5(2):021033, 2015.

\bibitem{Buxton19}
Samuel~J Buxton, David Quigley, and Scott Habershon.
\newblock The role of nuclear quantum effects in the relative stability of
  hexagonal and cubic ice.
\newblock {\em The Journal of Chemical Physics}, 151(14):144503, 2019.

\bibitem{Habershon09}
Scott Habershon, Thomas~E Markland, and David~E Manolopoulos.
\newblock Competing quantum effects in the dynamics of a flexible water model.
\newblock {\em The journal of chemical physics}, 131(2):024501, 2009.

\bibitem{GitHubRepo}
Pablo Piaggi.
\newblock https://github.com/pablopiaggi/phaseequilibriumwatericescan, 2021.

\end{thebibliography}

\end{document}